\documentclass[acmsmall]{acmart}
\AtBeginDocument{%
  }

\setcopyright{acmlicensed}
\copyrightyear{2018}
\acmYear{2018}
\acmDOI{XXXXXXX.XXXXXXX}
\acmConference[Conference acronym 'XX]{Make sure to enter the correct
  conference title from your rights confirmation email}{June 03--05,
  2018}{Woodstock, NY}
\acmISBN{978-1-4503-XXXX-X/2018/06}

\usepackage{listings}
\usepackage{xcolor} % 用于颜色定义

\definecolor{codebg}{rgb}{0.95,0.95,0.95} % 背景色
\definecolor{keyword}{rgb}{0,0,0.8}       % 关键字颜色
\definecolor{comment}{rgb}{0,0.6,0}      % 注释颜色

\lstdefinestyle{sqlstyle}{
	language=SQL,                        % 语言设为 SQL
	backgroundcolor=\color{codebg},      % 背景色
	basicstyle=\ttfamily\small,          % 基本字体
	keywordstyle=\color{keyword},        % 关键字颜色
	commentstyle=\color{comment},        % 注释颜色
	stringstyle=\color{red},             % 字符串颜色
	breaklines=true,                     % 自动换行
	frame=single,                        % 边框样式
	rulecolor=\color{gray},              % 边框颜色
	tabsize=2,                           % 缩进大小
	showstringspaces=false,              % 不显示字符串中的空格
	numbers=left,                        % 显示行号
	numberstyle=\tiny\color{gray},       % 行号样式
	captionpos=b                         % 标题位置（底部）
}

\usepackage{graphicx}
\usepackage{tabularx}  % 自动调整列宽
\usepackage{booktabs}  % 美化表格线
\usepackage{lipsum}    % 示例文本
\usepackage{algorithmic}
\usepackage{algorithm}
\usepackage{amsmath}

\usepackage{multirow}
\usepackage{caption}
\usepackage[inline]{enumitem}
\usepackage{enumitem} 
\usepackage{marvosym} 

\newcommand{\toolname}{FuzzySQL}

\usepackage{pifont}     % 钩和叉符号

\usepackage{etoolbox}
\AtBeginEnvironment{verbatim}{\vspace{-0.3em}}
\AtEndEnvironment{verbatim}{\vspace{-0.3em}}

\usepackage{threeparttable}

\newcommand{\frYes}{$\bullet$} % 全黑圆点
\newcommand{\frNo}{$\circ$}    % 空心圆

\usepackage{array}
\usepackage[most]{tcolorbox} 

\newtcolorbox{promptbox}[1]{%
	enhanced,
	colback=white,                 % 正文背景
	colframe=gray!60!black,        % 边框颜色
	boxrule=1.0pt,                 % 边框粗细
	arc=8pt,                       % 圆角
	outer arc=8pt,
	left=3.5mm,right=3.5mm,          % 内边距
	top=3mm,bottom=2.5mm,
	title={#1},                    % 标题
	fonttitle=\bfseries\Large,     % 标题字体
	coltitle=white,                % 标题文字颜色
	colbacktitle=gray!60!black,    % 标题栏背景
	toptitle=2mm,
	bottomtitle=2mm,
	titlerule=0mm,
	drop shadow=black!25,
}

\begin{document}

%%
%% The "title" command has an optional parameter,
%% allowing the author to define a "short title" to be used in page headers.
\title{\toolname: Uncovering Hidden Vulnerabilities in DBMS Special Features with LLM-Driven Fuzzing}

%%
%% The "author" command and its associated commands are used to define
%% the authors and their affiliations.
%% Of note is the shared affiliation of the first two authors, and the
%% "authornote" and "authornotemark" commands
%% used to denote shared contribution to the research.
\author{Yongxin Chen}
\affiliation{%
	\institution{National University of Defense Technology}
	\city{Changsha}
	\country{China}
}
\email{yongxinchen\_cx@nudt.edu.cn}

% 2 Zhiyuan Jiang (NUDT) - corresponding
\author{Zhiyuan Jiang}
\authornote{Zhiyuan Jiang and Yongjun Wang are corresponding authors.}
\affiliation{%
	\institution{National University of Defense Technology}
	\city{Changsha}
	\country{China}
}
\email{jzy@nudt.edu.cn}

% 3 Chao Zhang (Tsinghua)
\author{Chao Zhang}
\affiliation{%
	\institution{Tsinghua University}
	\city{Beijing}
	\country{China}
}
\email{chaoz@tsinghua.edu.cn}

% 4 Haoran Xu (NUDT)
\author{Haoran Xu}
\affiliation{%
	\institution{National University of Defense Technology}
	\city{Changsha}
	\country{China}
}
\email{xuhaoran12@nudt.edu.cn}

% 5 Shenglin Xu (NUDT)
\author{Shenglin Xu}
\affiliation{%
	\institution{National University of Defense Technology}
	\city{Changsha}
	\country{China}
}
\email{xushenglin@nudt.edu.cn}

% 6 Jianping Tang (HNU)
\author{Jianping Tang}
\affiliation{%
	\institution{Hunan University}
	\city{Changsha}
	\country{China}
}
\email{tjp@hnu.edu.cn}

% 7 Zheming Li (Tsinghua)
\author{Zheming Li}
\affiliation{%
	\institution{Tsinghua University}
	\city{Beijing}
	\country{China}
}
\email{lizm20@mails.tsinghua.edu.cn}

% 8 Peidai Xie (NUDT)
\author{Peidai Xie}
\affiliation{%
	\institution{National University of Defense Technology}
	\city{Changsha}
	\country{China}
}
\email{xpd2002@126.com}

% 9 Yongjun Wang (NUDT) - corresponding
\author{Yongjun Wang}
\authornotemark[1]
\affiliation{%
	\institution{National University of Defense Technology}
	\city{Changsha}
	\country{China}
}
\email{wangyongjun@nudt.edu.cn}

%%
%% By default, the full list of authors will be used in the page
%% headers. Often, this list is too long, and will overlap
%% other information printed in the page headers. This command allows
%% the author to define a more concise list
%% of authors' names for this purpose.
\renewcommand{\shortauthors}{Chen et al.}

%%
%% The abstract is a short summary of the work to be presented in the
%% article.
\begin{abstract}
	Traditional database fuzzing techniques primarily focus on syntactic correctness and general SQL structures, leaving critical yet obscure DBMS features,  
	such as system-level modes
	(e.g., GTID), 
	programmatic constructs
	(e.g., PROCEDURE), 
	advanced process commands
	(e.g., KILL), 
	largely underexplored. 
	Although rarely triggered by typical inputs, these features can lead to severe crashes or security issues when executed under edge-case conditions.
	In this paper, we present \toolname, a novel LLM-powered adaptive fuzzing framework designed to uncover subtle vulnerabilities in DBMS special features. 
	\toolname\ combines grammar-guided SQL generation with logic-shifting progressive mutation, a novel technique that explores alternative control paths by negating conditions and restructuring execution logic, synthesizing structurally and semantically diverse test cases.
	To further ensure deeper execution coverage of the back end, \toolname\ employs a hybrid error repair pipeline that unifies rule-based patching with LLM-driven semantic repair, enabling automatic correction of syntactic and context-sensitive failures.
	We evaluate \toolname\ across multiple DBMSs, including MySQL, MariaDB, SQLite, PostgreSQL and Clickhouse, uncovering 64 vulnerabilities, 27 of which are tied to under-tested DBMS special features. As of this writing, 60 cases have been confirmed with 9 assigned CVE identifiers, 31 already fixed by vendors, and additional vulnerabilities scheduled to be patched in upcoming releases.
	Our results highlight the limitations of conventional fuzzers in semantic feature coverage and demonstrate the potential of LLM-based fuzzing to discover deeply hidden bugs in complex database systems.
\end{abstract}

%%
%% The code below is generated by the tool at http://dl.acm.org/ccs.cfm.
%% Please copy and paste the code instead of the example below.
%%
\begin{CCSXML}
	<ccs2012>
	<concept>
	<concept_id>10002978.10003018</concept_id>
	<concept_desc>Security and privacy~Database and storage security</concept_desc>
	<concept_significance>500</concept_significance>
	</concept>
	<concept>
	<concept_id>10002978.10003022</concept_id>
	<concept_desc>Security and privacy~Software and application security</concept_desc>
	<concept_significance>500</concept_significance>
	</concept>
	</ccs2012>
\end{CCSXML}
\ccsdesc[500]{Security and privacy~Database and storage security}
\ccsdesc[500]{Security and privacy~Software and application security}

%%
%% Keywords. The author(s) should pick words that accurately describe
%% the work being presented. Separate the keywords with commas.
\keywords{Database security, Fuzzing, AI for security,  Language validity}

\received{20 February 2007}
\received[revised]{12 March 2009}
\received[accepted]{5 June 2009}

%%
%% This command processes the author and affiliation and title
%% information and builds the first part of the formatted document.
\maketitle

\section{Introduction}
Modern database management systems (DBMSs) sit at the core of software infrastructure that is critical to security and reliability, where subtle logic flaws can lead to persistent crashes, state corruption, and severe service disruption. 
As a result, fuzzing has emerged as an effective technique for uncovering DBMS bugs by generating SQL statements and executing them against target systems.
However, most existing DBMS fuzzers~\cite{25sqlsmith,27zhong2020squirrel,08rigger2020testing,22liang2022detecting,18jiang2024detecting} predominantly focus on general-purpose statement structures (e.g., \texttt{SELECT}), covering only a limited portion of the DBMS grammar. 
Recent efforts such as BUZZBEE~\cite{47yang2024towards} and POLYGLOT~\cite{69chen2021one} emphasize cross-target generalizability, rather than systematically exercising DBMS-specific features such as read-only modes or \texttt{GTID} control.

DBMS special features are vendor-specific extensions that deviate from standard SQL specifications and, although invoked less frequently, are deeply embedded in core execution paths and can trigger severe system failures under edge conditions.
For example, Listing~\ref{lst:motivation_example} shows that a seemingly benign combination of statements, such as \texttt{UPDATE HISTOGRAM} and \texttt{RESET} within a read-only transaction, can lead to a full-system crash—behaviors that are rarely exercised by conventional fuzzers.
Existing DBMS fuzzers have explored SQL generation and mutation from multiple perspectives, including structural complexity~\cite{27zhong2020squirrel,26jiang2023dynsql}, sequence-level interactions~\cite{29fu2022griffin,30liang2023sequence}, and oracle-guided logic testing~\cite{11rigger2020detecting,22liang2022detecting,18jiang2024detecting}. 
Despite these advances, they share a common limitation: none have made effective attempts to systematically explore feature-specific logic.
At a fundamental level, this limitation stems from the mismatch between feature-specific DBMS behaviors and the design assumptions of existing fuzzers, and manifests in several aspects.

Concretely, we identify three key factors that hinder existing fuzzers from effectively exploring feature-specific DBMS logic.
First, traditional fuzzers lack support for feature-specific SQL constructs that are often tied to internal DBMS mechanisms. 
These constructs are not only vendor-specific but also version-dependent, making rule-based modeling costly and slow to adapt.
Second, semantic modeling remains a fundamental challenge. Without understanding complex schema dependencies and execution context, traditional tools struggle to synthesize auxiliary objects  (e.g., \texttt{PROCEDURE}) or establish the preconditions needed to exercise deep execution paths, leading to low coverage of feature-specific logic.
Third, existing fuzzers typically lack global state tracking and replay mechanisms, making it difficult to identify compound failures that arise only through interactions across multiple statements or test cases. For example, the underlying bug demonstrated by Listing~\ref{lst:motivation_example} was triggered by a latent interaction among statements from four distinct test cases.

\vspace{0.2cm}
\begin{minipage}{0.97\textwidth}
	\begin{lstlisting}[style=sqlstyle, caption={Motivational example of a MySQL vulnerability related to DBMS special features.}, label=lst:motivation_example]
		CREATE TABLE t0 (xxx);
		ALTER DATABASE test_db READ ONLY = 1;
		ANALYZE TABLE `t0` UPDATE HISTOGRAM ON v1;
		RESET BINARY LOGS AND GTIDS;
		INSERT INTO t0 VALUES (xxx);
	\end{lstlisting}
\end{minipage}

To address these limitations, we present \toolname, a novel adaptive fuzzing framework driven by LLMs and designed to systematically uncover vulnerabilities in the under-tested feature space of modern DBMSs. \toolname\ introduces the following key innovations.
First, \toolname\ targets feature-specific SQL constructs by combining grammar-guided template expansion with LLM-powered instantiation (Section~\ref{sec:Generation}), enabling systematic coverage of vendor-specific and version-dependent DBMS features that are difficult to model using fixed rules.
Second, to overcome the challenges of semantic modeling, \toolname\ performs semantic exploration through a unified design that integrates progressive mutation with automated error repair (Sections~\ref{sec:Mutation} and~\ref{sec:Repairation}).
Specifically, it prioritizes semantic diversity by recombining logic fragments across seed programs and applying conditional logic shifts, while automatically repairing high-value test cases that fail due to missing objects or unmet semantic preconditions.
This design enables \toolname\ to construct semantically valid and diverse SQL sequences that reach deep, feature-dependent execution paths.
Finally, \toolname\ maintains the global execution context throughout fuzzing and performs replay-based analysis to identify and isolate compound failures that arise only under specific internal DBMS states (Section~\ref{sec:Validation}).
This paper makes the following key contributions:

\begin{itemize}[leftmargin=1.8em]
	\item \textbf{Uncovering a critical blind spot in DBMS fuzzing.}
	We demonstrate that conventional fuzzers typically neglect the special features of DBMSs, such as \texttt{GTID} handling, which can lead to severe failures, but are rarely triggered by general-purpose test cases.
	\item \begin{sloppypar} \textbf{Enabling deep semantic exploration through LLM-driven fuzzing.}
		We present \toolname, an adaptive fuzzing framework that combines template-guided generation, context-aware mutation, and a hybrid error-repair pipeline. This design enables the synthesis of structurally valid and semantically rich SQL statement sequences that activate deeper DBMS control paths.
	\end{sloppypar}
	\item \textbf{Comprehensive feature coverage via execution state maintenance and replay.}
	\toolname\ flexibly adapts to diverse DBMS dialects and feature sets and maintains the global execution context during fuzzing and crash validation. This allows it to detect feature-related bugs that depend on internal state transitions and diverse execution environments.
	\item 
	\begin{sloppypar}\textbf{Discovering real-world vulnerabilities across multiple DBMSs.}
		We evaluate \toolname\ on MySQL, MariaDB, SQLite, and ClickHouse, uncovering 64 vulnerabilities, of which 60 have been confirmed by developers and 27 are associated with under-tested DBMS features.
		These results highlight \toolname's ability to expose semantic and feature-specific vulnerabilities consistently missed by existing approaches.
	\end{sloppypar}
\end{itemize}
% Upon acceptance, we will release the prototype implementation of \toolname\ as open-source to facilitate reproducibility and further research.
% We provide an anonymized artifact repository  to support reproducibility, as described in the Open Science appendix.

\section{Motivation}

Modern DBMSs implement a wide range of system-specific features that go far beyond standard SQL support, introducing complex internal mechanisms and stateful behaviors. These under-tested features often expose latent vulnerabilities that are difficult to uncover through traditional fuzzing techniques. In this section, we first highlight the limitations of existing fuzzers in exercising such features through a motivating example. We then discuss why previous approaches fail to trigger these bugs and explore the potential of LLMs to address the associated challenges.

\subsection{Under-Tested Features in Modern DBMSs}

Traditional database fuzzers mainly focus on general-purpose SQL statements such as \texttt{SELECT}, \texttt{INSERT}, and \texttt{CREATE TABLE}. However, modern DBMSs expose a much broader attack surface through SQL-accessible system features that manipulate internal metadata, runtime modes, and persistent execution state. Listing~\ref{lst:motivation_example} shows a minimal SQL sequence that triggers a persistent crash in MySQL by exercising such under-tested feature interactions.

In this paper, we use \emph{under-tested DBMS features} to refer to functionalities that are vendor-specific, version-evolving, or operationally uncommon, yet directly affect core DBMS subsystems. Representative categories include configuration and state toggles (e.g., \texttt{READ ONLY}), replication and logging controls (e.g., \texttt{BINARY LOGS} and \texttt{GTID}), and optimizer or statistics maintenance operations (e.g., \texttt{UPDATE HISTOGRAM}). Thus, the challenge is broader than covering rare syntax alone: these features typically require non-trivial semantic preconditions and often interact with sensitive internal subsystems such as metadata management, replication logic, and statistics maintenance.

The example in Listing~\ref{lst:motivation_example} illustrates this challenge. Each statement is syntactically valid and semantically reasonable in isolation, yet their particular combination exposes a subtle vulnerability in MySQL’s metadata lock management. Specifically, toggling \texttt{READ ONLY} mode and updating table histograms affect metadata locks maintained by the dictionary subsystem, while resetting binary logs modifies session-level replication state. When followed by a regular \texttt{INSERT}, these accumulated effects violate internal lock invariants and trigger an assertion failure in the metadata lock checker.
More importantly, the bug is state-dependent and persistent. Once triggered, it corrupts internal state in a way that survives subsequent executions: even after restarting the server, later queries may still crash immediately. This example shows that feature-related DBMS bugs are hard to expose not simply because such statements are less frequently generated, but because they require the fuzzer to jointly cover feature-specific constructs, satisfy their semantic preconditions, and preserve the execution history in which harmful interactions emerge.

\subsection{Why Existing Fuzzers Miss These Bugs}
Despite progress in DBMS fuzzing, most existing tools remain centered on crafting syntactically valid DML and DDL statements, with limited awareness of feature-specific SQL constructs. Moreover, such features vary significantly across DBMSs and evolve rapidly between releases. For example, \texttt{RESET BINARY LOGS AND GTIDS} in Listing~\ref{lst:motivation_example} is a recent feature introduced in MySQL~8.4 to replace the legacy \texttt{RESET MASTER}, making manual rule-based support both costly and prone to lag behind DBMS evolution.

\begin{figure*}[htbp]
	\setlength{\abovecaptionskip}{3pt} 
	\centering
	\includegraphics[width=.99\textwidth]{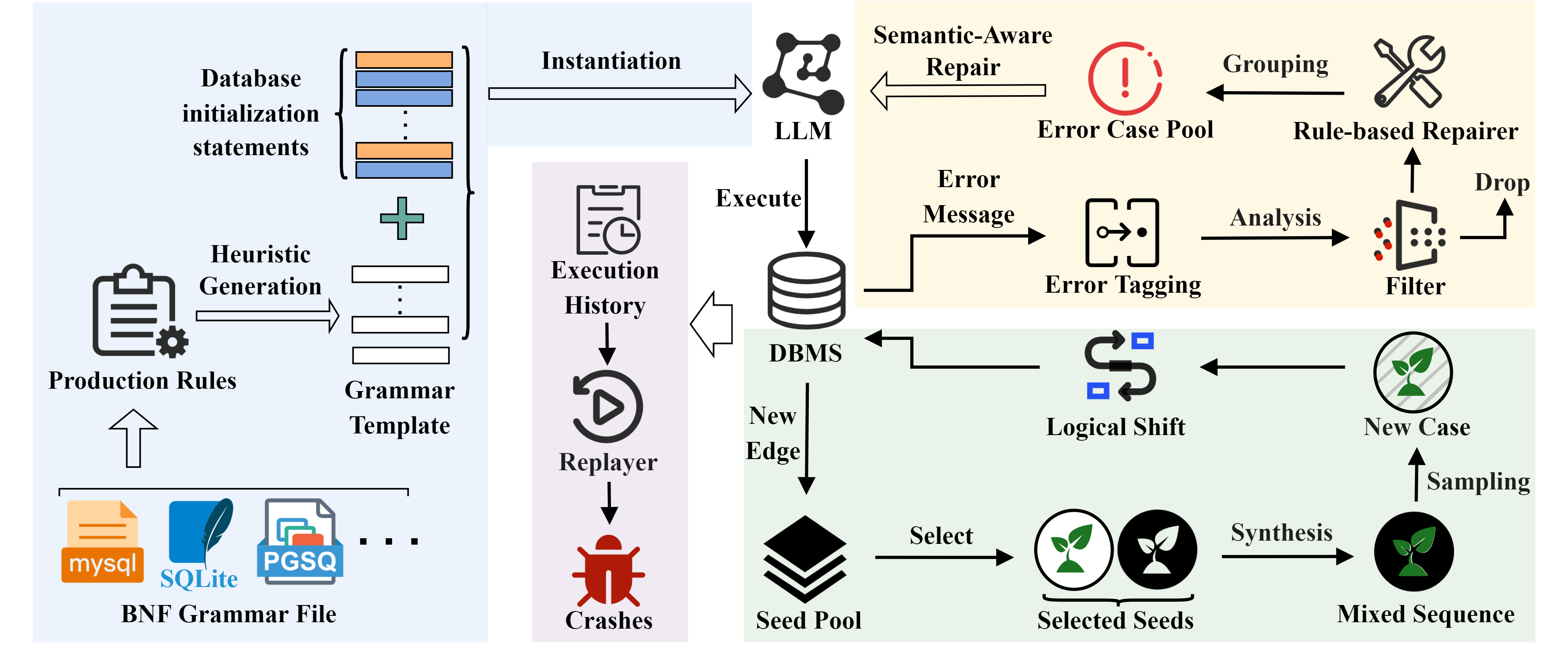} 
	\caption{Overview of FuzzySQL.} 
	\label{fig:FuzzySQL} 
\end{figure*}

In addition, prior tools typically lack global state awareness. Without tracking execution context or supporting replay, they may overlook or misclassify state-dependent failures as false positives. In our investigation, the vulnerability shown in Listing~\ref{lst:motivation_example} could not be triggered by a single test case; instead, it was reconstructed from four independent inputs identified among thousands of PoCs. These inputs were discovered through sequential replay and input minimization, as there was no semantic guidance suggesting that \texttt{READ ONLY} toggles and \texttt{RESET} commands should be combined. This limitation underscores that current fuzzers are fundamentally inadequate for exposing deep, feature-related vulnerabilities.

\subsection{LLM: Opportunities and Challenges}
LLMs provide an effective complement to traditional DBMS fuzzing because they can reason about SQL at the semantic level rather than only at the syntactic level. This capability is particularly useful for exercising under-tested DBMS features, whose correct invocation often depends on implicit object dependencies, valid statement ordering, and feature-specific execution context. In practice, LLMs can generate coherent multi-statement SQL sequences, infer missing prerequisites, and synthesize auxiliary objects or definitions (e.g., creating a \texttt{PROCEDURE} before issuing a \texttt{CALL}). By doing so, they reduce the need for manually crafted feature-specific rules and improve the ability of the fuzzer to reach semantically constrained execution paths.

However, using LLMs for fuzzing also introduces challenges. Since these models are trained and aligned primarily on well-formed examples, they are naturally inclined to produce syntactically correct, semantically plausible, and conventionally safe outputs. This tendency is useful for standard code generation, but less ideal for fuzzing, where valuable test cases often arise from unusual feature combinations, edge-case dependencies, or execution contexts that depart from typical usage. As a result, LLMs must be guided carefully so that they reason over schema context and execution history instead of defaulting to generic low-risk completions.

A related issue is hallucination, where the model generates content that does not fully match the actual DBMS dialect, schema, or runtime context. Although such deviations are usually undesirable, they can occasionally introduce exploratory diversity beyond normative SQL patterns and thereby help expose unexpected feature interactions or corner-case failures. For this reason, \toolname\ does not treat LLM output as inherently trustworthy or inherently harmful; instead, it uses downstream execution feedback and repair to retain useful exploratory behaviors while filtering out low-value failures. We further illustrate this phenomenon through case studies in Section~\ref{subsec:Casestudy}.

\subsection{Goals of \toolname}
Based on the motivations described above, we design \toolname\ as an LLM-powered adaptive fuzzing framework that aims to cover the under-tested feature space of DBMS.
\toolname\ is designed to:

\begin{enumerate}[leftmargin=2.2em]
	\renewcommand{\labelenumi}{\roman{enumi}.} % 小写罗马数字 + 点号
	\item Flexibly adapt to diverse DBMS dialects and system-specific features through grammar-based template expansion and LLM-guided semantic instantiation.
	\item Generate syntactically valid and semantically meaningful SQL sequences that invoke diverse internal behaviors.
	\item Maintain global execution context and support replay-based PoC validation to detect and isolate persistent, state-dependent bugs rooted in non-trivial control flows.
\end{enumerate}

Overall, \toolname\ aims to shift DBMS fuzzing beyond purely syntax-driven generation or random mutation toward a semantically guided workflow that explores both deeper program logic and broader system functionality.
More importantly, \toolname\ is not merely a general-purpose LLM-assisted SQL fuzzer. Its design explicitly addresses three obstacles that disproportionately hinder the discovery of feature-related DBMS bugs: (1) insufficient coverage of vendor- and version-specific feature constructs; (2) unmet semantic preconditions that prevent such features from being exercised effectively; and (3) hidden state interactions that span multiple statements or executions. Accordingly, \toolname\ combines feature-oriented grammar expansion and LLM instantiation, semantics-aware mutation and repair, and replay-guided validation under persistent execution context.

\section{Methodology}
To uncover deep and stateful bugs in modern DBMSs, \toolname\ adopts a unified fuzzing pipeline that integrates structured generation, semantic mutation, automated repair, and replay-based validation, as illustrated in Figure~\ref{fig:FuzzySQL}. Rather than generating isolated SQL statements, \toolname\ continuously evolves multi-statement programs under a persistent execution context, enabling exploration of feature-specific and state-dependent behaviors.
Fuzzing begins with grammar-guided SQL template expansion, followed by LLM-based instantiation under the current schema state to produce executable and semantically meaningful inputs. Executed test cases that yield new coverage are retained and further evolved via progressive mutation, which recombines statement sequences and applies logic-level transformations to explore alternative control paths. When execution fails due to syntactic or semantic inconsistencies, \toolname\ applies a multi-stage repair pipeline to salvage high-value inputs instead of discarding them. Finally, detected crashes are validated through replay-guided execution and minimized to concise, reproducible PoCs that capture state-dependent failures.
Each stage is designed to address a distinct obstacle in uncovering feature-related DBMS bugs: grammar-guided generation expands the reachable feature space, mutation and repair satisfy non-trivial semantic preconditions, and replay-guided validation exposes failures whose triggers are distributed across persistent execution history.

\subsection{Grammar-Guided SQL Generation}
\label{sec:Generation}
To generate diverse and structurally valid SQL inputs, \toolname\ employs a two-phase grammar-driven generation process. First, it performs a recursive expansion of production rules extracted from the DBMS grammar definitions. Second, it uses LLM to instantiate concrete SQL statements guided by both the expanded templates and the contextual schema state.

\begin{figure}[htbp]
	\setlength{\abovecaptionskip}{2pt} 
	\centering
	\includegraphics[width=0.6\textwidth]{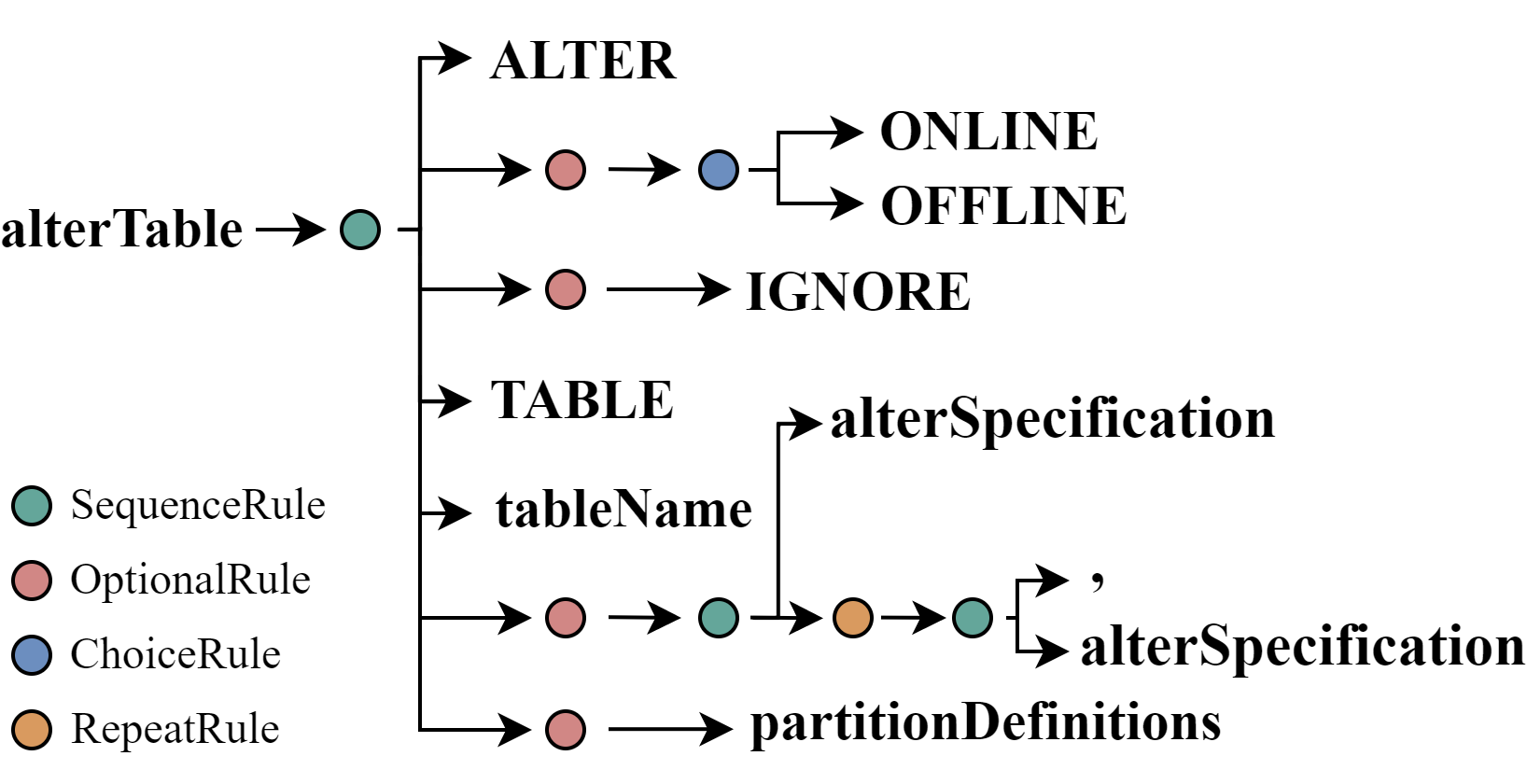}
	\caption{CFG rules of the `alterTable' statement.}
	\label{fig:alterTable}
\end{figure}

\subsubsection{Grammar Template Expansion}
To systematically cover the SQL grammar space, 
\toolname\ first parses the target DBMS's ANTLR4 grammar specification to extract production rules.
We abstract all rules into four categories to enable controlled and interpretable expansion, as illustrated in Figure~\ref{fig:alterTable}:

\begin{itemize}
	\item \textbf{SequenceRule:} A sequence of grammar symbols to be expanded in order.
	\item \textbf{OptionalRule:} Grammar elements that may be omitted or included (e.g., \texttt{IGNORE}).
	\item \textbf{ChoiceRule:} Mutually exclusive options, where one is selected from a set (e.g., \texttt{ONLINE} vs. \texttt{OFFLINE}).
	\item \textbf{RepeatRule:} Patterns that can be repeated one or more times (e.g., comma-separated lists of {\ttfamily alter\-Spec\-i\-fi\-ca\-tion}).
\end{itemize}

\begin{minipage}{0.97\textwidth}
	\begin{lstlisting}[style=sqlstyle, caption={SQL templates derived from the alterTable rule.}, label=lst:altertable,
		breaklines=true,      
		breakatwhitespace=false,
		breakautoindent=true,
		literate=
		{ter}{{ter}}3
		]
		alterTable -> ALTER ONLINE TABLE [tableName] [alterSpecification] [partitionDefinitions]
		alterTable -> ALTER TABLE [tableName] [alter-Specification],[alterSpecification]
		
	\end{lstlisting}
\end{minipage}

Based on this classification, we scan and normalize the grammar file, enabling heuristic control during recursive expansion. 
Then, \toolname\ constructs recursive derivation trees rooted at top-level statement nonterminals.
Two safeguards are applied to ensure diversity and manageability:
\begin{enumerate*}[label=(\roman*), itemjoin={{; }}]
	\item A configurable maximum recursion depth to bound nested grammar derivations
	\item Rule-specific usage quotas to prevent structural redundancy (e.g., excessive parentheses or nested subqueries).
\end{enumerate*}
The final output is a collection of skeletal SQL templates—syntax-valid but uninstantiated statement blueprints. 
Listing \ref{lst:altertable} shows examples of grammar templates expanded from the \texttt{alterTable} rule. 
Both templates can be further expanded through the \texttt{alterSpecification} rule, as long as the depth and quota limits are not exceeded.
Grammar templates form the structural backbone of FuzzySQL's seed generation process, ready to be instantiated with real schema and data context by LLMs in the next phase.

\begin{sloppypar}
	\subsubsection{LLM-Powered Fuzzy Instantiation}
	Given a grammar-expanded template, \toolname\ uses a LLM to instantiate the abstract statement into a complete executable SQL query. To guide this process, we maintain a dynamic schema context—populated via randomly generated \texttt{CREATE TABLE} and \texttt{INSERT} statements, which includes table names, column types and relational structures. These initialization statements not only establish a valid database environment with rich type coverage (e.g., \texttt{INT}, \texttt{ENUM}, \texttt{GEOMETRY}), but also provide the semantic context necessary for the LLM to generate coherent and meaningful SQL logic.
\end{sloppypar}

For example, when instantiating a template involving a \texttt{JOIN} or subquery, the LLM leverages both the schema and the grammatical scaffold to synthesize a valid and meaningful test case. Unlike traditional fuzzers, which randomly inject literals or select columns uniformly, \toolname\ is context-aware and schema-consistent. As a result, \toolname\ can generate a wide variety of semantically correct and structurally diverse SQL queries ranging from simple data manipulation to complex administrative commands while ensuring that they are executable in the current DBMS state.

\setlength{\textfloatsep}{5pt}
\begin{algorithm}[!t]
	\caption{Logic-Shifting Progressive Mutation}
	\label{alg:mutation}
	\begin{algorithmic}[1]
		\STATE \textbf{Input:} Two SQL test cases $\mathcal{T}_1$, $\mathcal{T}_2$
		\STATE \textbf{Output:} Mutated test case $\mathcal{T}^*$
		\STATE Extract schema fragments $\mathcal{S}_1$, $\mathcal{S}_2$ and operations $\mathcal{O}_1$, $\mathcal{O}_2$
		\STATE Initialize empty schema set $\mathcal{S} \gets \emptyset$
		\FOR{each table $t$ in $\mathcal{S}_1 \cup \mathcal{S}_2$}
		\IF{$t$ exists in both $\mathcal{S}_1$ and $\mathcal{S}_2$}
		\STATE Randomly select one version of $t$ from $\mathcal{S}_1$ or $\mathcal{S}_2$
		\ELSE
		\STATE Use the available definition of $t$
		\ENDIF
		\STATE Append chosen schema statements of $t$ into $\mathcal{S}$
		\ENDFOR
		\STATE Initialize empty operation sequence $\mathcal{O} \gets EmptyList$
		\STATE Let $i, j$ be indices for $\mathcal{O}_1$, $\mathcal{O}_2$
		\WHILE{$i < |\mathcal{O}_1|$ or $j < |\mathcal{O}_2|$}
		\IF{$i < |\mathcal{O}_1|$ and ($j \geq |\mathcal{O}_2|$ or random $< 0.5$)}
		\STATE Append $\mathcal{O}_1[i]$ to $\mathcal{O}$; $i \gets i + 1$
		\ELSE
		\STATE Append $\mathcal{O}_2[j]$ to $\mathcal{O}$; $j \gets j + 1$
		\ENDIF
		\ENDWHILE
		\FOR{each statement $s$ in $\mathcal{O}$}
		\STATE Sample a drop probability $p_{\text{drop}} \sim \mathcal{U}(0.2, 0.4)$
		\IF{$random() < p_{\text{drop}}$}
		\STATE Drop $s$
		\ENDIF
		\ENDFOR
		\FOR{each statement $s$ in $\mathcal{O}$}
		\STATE Apply conditional rewrites $(k \rightarrow k')$:
		\STATE \quad if $k$ is JOIN, randomly replace with supported JOIN;
		\STATE \quad else replace with $k'$
		\ENDFOR
		\STATE \textbf{Return:} $\mathcal{T}^* = \mathcal{S}$ $\cup$ $\mathcal{O}$
		
	\end{algorithmic}
	
\end{algorithm}

\subsection{Logic-Shifting Progressive Mutation}
\label{sec:Mutation}

While grammar-based generation ensures structural diversity, mutation plays a crucial role in evolving test cases toward deeper execution paths. 
\toolname\ introduces context reconstruction that operates at the SQL sequence level, incorporating a logic-shifting progressive mutation strategy to maximize semantic diversity and minimize redundancy.
This design avoids complex parsing and rigid mutation rules, enabling efficient discovery of diverse behaviors across DBMS dialects with minimal seed input requirements.
The overall process is illustrated in Algorithm~\ref{alg:mutation}.

\subsubsection{Sequence Crossover Synthesis}
To generate new test cases, \toolname\ randomly selects a pair of seed SQL statement sequences from the seed pool and performs schema-aware crossover synthesis. The process follows 3 steps:
\textbf{Schema Unification.} As shown in lines 3–11 of Algorithm~\ref{alg:mutation}, we first extract schema initialization statements (e.g., \texttt{CREATE TABLE}, \texttt{INSERT INTO}) from both $T_1$ and $T_2$, denoted as $S_1$ and $S_2$. For each table identifier $t$ appearing in $S_1 \cup S_2$, we resolve potential conflicts: if $t$ is defined in both seeds, one version is randomly chosen to ensure consistency and eliminate redundancy. The unified schema statements are appended to form a consistent schema set $S$.

\textbf{Sequence Crossover.} 
We extract non-initialization SQL statements (e.g., \texttt{SELECT}, \texttt{ANALYZE}, \texttt{KILL}) from $T_1$ and $T_2$ to form operation sequences $\mathcal{O}_1$ and $\mathcal{O}_2$. As shown in Lines~13--21 of Algorithm~\ref{alg:mutation}, these sequences are probabilistically interleaved to produce a new sequence $\mathcal{O}$ that preserves partial order while enabling cross-seed logic recombination. This allows \toolname\ to synthesize test cases that blend behaviorally diverse fragments and explore context-dependent interactions.
Sequence crossover may introduce semantic inconsistencies, such as mismatched table or column references. To reduce identifier conflicts, we adopt a unified naming convention during generation (e.g., \texttt{t}, \texttt{c} with numeric suffixes). 
Moreover, syntactic or semantic errors introduced during mutation are not fatal—they are addressed in the subsequent error repair stage.

\textbf{Size and Diversity Control.}
To avoid overly large or redundant test cases, we perform probabilistic dropping of operations (lines 22–27). For each statement $s$ in $\mathcal{O}$, a drop probability $p_\text{drop}$ is sampled from a uniform distribution $\mathcal{U}$, and $s$ is discarded with that probability.
This strategy ensures that the generated test cases remain compact, computationally efficient, and support progressive exploration of novel context combinations.
The resulting synthetic case blends the execution contexts and logic flows from both parents, increasing the likelihood of triggering corner-case bugs that require complex state transitions.

\subsubsection{Conditional Logic Shift}
After crossover synthesis, \toolname\ applies a second-stage mutation phase that targets the logical structure of SQL conditions. This logic-shifting transformation rewrites common predicates and joins patterns to explore alternative semantic paths while preserving syntactic validity. The process is implemented via rule-driven string rewrites, requiring no schema-specific reasoning or AST instrumentation, and generalizes well across SQL dialects.
This logic-level rewriting corresponds to lines 28–32 in Algorithm~\ref{alg:mutation}, where each statement $s$ in the filtered operation sequence is examined for conditional rewrite opportunities. Specifically, \toolname\ maintains a rewrite rule set $\{k \rightarrow k'\}$, and applies them conditionally based on statement type.
We organize these transformations into two primary categories:

\begin{enumerate}[label=(\roman*), itemjoin={{; }}]
	\renewcommand{\labelenumi}{\roman{enumi}.} 
	\item \textbf{Predicate mutations}, which rewrite conditional operators (e.g., \texttt{=} vs.\ \texttt{!=}, \texttt{IN} vs.\ \texttt{NOT IN}, \texttt{IS NULL} vs.\ \texttt{IS NOT NULL}, or ordering directions) in \texttt{WHERE}, \texttt{HAVING}, and \texttt{ORDER BY} clauses. These lightweight changes preserve syntactic structure while enabling exploration of complementary execution paths.
	\item \textbf{\texttt{JOIN} rewrites}, which substitute join variants (e.g., \texttt{INNER}, \texttt{LEFT}, or \texttt{CROSS JOIN}) to perturb query structure and query-plan generation, potentially exposing logic differences in optimization or error-handling stages.
\end{enumerate}

In contrast to AST-based fuzzers such as Squirrel~\cite{27zhong2020squirrel} that rely on heavyweight intermediate representations, our mutation layer is minimal and DBMS-agnostic, allowing efficient fuzzing over LLM-generated SQL with low engineering cost. To maintain cross-system compatibility, \toolname\ includes dialect-aware filtering: for instance, disabling \texttt{FULL JOIN} in SQLite or adapting join syntax for ClickHouse. This ensures that semantic variation remains within valid syntactic boundaries for each target engine.

\subsection{Automated Error Repair}
\label{sec:Repairation}
Instantiating complex grammar templates into fully valid SQL statements in a single pass is inherently difficult, particularly for under-tested features such as \texttt{PROCEDURE}. In addition, random mutations often introduce contextual inconsistencies, producing test cases that are syntactically invalid yet still encode meaningful execution intent. Rather than relying on one-shot generation to produce high-quality inputs, \toolname\ treats test case construction as a refinement process and introduces a progressive error recovery pipeline. By repairing and completing partially valid SQL programs, the system systematically elevates test case quality and enables exploration of deeper, feature-dependent execution paths that are difficult to reach through direct generation alone. Embracing imperfection in exchange for broader exploration, \toolname\ aligns naturally with the fundamental philosophy of fuzzing.

\begin{table*}[htbp]
	\setlength{\abovecaptionskip}{4pt} 
	\renewcommand{\arraystretch}{0.97}
	\centering
	\caption{SQL Error Handling Strategies}
	\label{tab:error_repair}
	\newcommand{\SAF}{\textbf{SAF}}
	\newcommand{\RBR}{\textbf{RBR}}
	\newcommand{\SAR}{\textbf{SAR}}
	\newcolumntype{C}[1]{>{\centering\arraybackslash}p{#1}}
	\begin{tabularx}{\textwidth}{ 
			>{\hsize=0.85\hsize}X
			>{\hsize=1.05\hsize}X
			>{\hsize=1.1\hsize}X C{0.6cm} C{0.6cm} C{0.6cm}}
		% \hline
		\bottomrule
		\textbf{Error Category} & \textbf{Definition} & \textbf{Repair Strategy} & \textbf{\SAF} & \textbf{\RBR} & \textbf{\SAR} \\
		% \hline
		\midrule
		Duplicate Definition  & A data object is defined more than once. & \raisebox{0.2ex}{--}  &  \checkmark & & \\
		\hline
		% \midrule
		Unsupported Feature & Uses a feature or syntax not supported by the system. & \raisebox{0.2ex}{--}  & \checkmark & & \\
		\hline
		Plugin / Component Errors  & Errors caused by failure or malfunction of plugins or components. & Reload, reconfigure, or replace the plugin/component. & & \checkmark & \\
		\hline
		Inappropriate Setting & A configuration setting is invalid or contextually incorrect. & Adjust to a valid or recommended setting. & & \checkmark & \\
		\hline
		Formattable Errors & 	Errors that can be resolved through formatting adjustments. & Apply predefined repair rules through SQL statement formatting analysis. & & \checkmark & \\
		\hline
		Invalid Object Reference & References a non-existent or illegal data object. & Generates a SQL statement template to create or define the object and inserts it in front. & & & \checkmark \\
		\hline
		Preconditions Missing  & 	Required preconditions for execution are not met. & Extract prerequisite requirements and build guiding prompts. & & & \checkmark \\
		\hline
		Incorrect Feature Usage  & Incorrect or improper usage of a feature. & Attach the error messages directly to the corresponding SQL statements. & & & \checkmark \\
		\hline
		Violate Constraints & Operation violates database constraints. & Attach the error messages directly to the corresponding SQL statements. & & & \checkmark \\
		
		% \hline
		\toprule
	\end{tabularx}
	
	\footnotesize
	\parbox{\textwidth}{\hfill 
		\textbf{\SAF}: Syntax-Aware Filtering
		\textbf{\RBR}: Rule-Based Repair \textbf{\SAR}: Semantic-Aware Repair \hfill}

\end{table*}

\subsubsection{Error Taxonomy}

To ensure the soundness of our repair strategy, we conduct a quantitative analysis of error messages returned by MySQL, a representative and widely deployed DBMS. From 12 hours of grammar-guided SQL generation, we collect 65,591 errors spanning 184 distinct types (out of 1,960 documented MySQL client error codes). Syntax-related errors account for 29.61\% of all messages, while the top 50 semantic error types contribute 68.44\%. 
These high-frequency semantic errors are manually analyzed and mapped to dedicated rule-based or intelligent semantic-aware repair strategies.

Following the same methodology, we perform 12-hour fuzzing sessions for other DBMSs and map the observed error patterns to the corresponding repair strategies. Although MySQL serves as the primary guide for designing our framework, \toolname\ maintains compatibility across systems by adapting repairs based on error semantics rather than relying on DBMS-specific rules. The SQL error handling strategies we summarized are shown in Table \ref{tab:error_repair}.

\subsubsection{Syntax-Aware Filtering}
Before invoking any repair strategy, \toolname\ first classifies execution errors according to the taxonomy introduced in Section~3.3.1. This step determines whether an error should be repaired or filtered out directly. In particular, \toolname\ focuses repair efforts on syntax and semantic errors that still indicate structurally or semantically promising SQL inputs, while excluding failures that are unlikely to contribute useful execution behaviors.

To this end, \toolname\ maintains a lightweight, DBMS-specific set of error patterns. Each pattern is defined by an error code, a regular-expression-style keyword matcher over the DBMS error message, and an associated action. When an error matches a filtering rule, \toolname\ labels it as \texttt{SAF} and discards the corresponding input from further repair. Typical examples include duplicate-definition errors (e.g., messages containing \texttt{already exists}) and unsupported-feature errors (e.g., \texttt{Method ... is not supported}). These failures usually indicate environmental conflicts or feature absence rather than repairable input defects, and thus provide little value for subsequent mutation or semantic correction.
For the remaining syntax-related failures, \toolname\ applies structural filtering to prioritize repair-worthy candidates. Specifically, statements with fewer than twenty tokens are more likely to have an abstract syntax tree (AST) depth of fewer than three; such statements are discarded as overly shallow and unlikely to exercise deep execution logic.

This filtering strategy allows \toolname\ to avoid wasting repair budget on low-value failures while preserving inputs that are more likely to trigger meaningful feature interactions or deeper internal behaviors. In practice, the goal of this stage is not compiler-grade diagnostic precision, but to narrow the repair scope sufficiently so that subsequent repair modules can focus on high-value candidates.

\subsubsection{Rule-Based Repair.}
After syntax- and error-aware filtering, \toolname\ applies a rule-based repair (RBR) layer to handle error cases that admit predictable, pattern-driven fixes. Similar to the filtering stage, this module is triggered by DBMS-specific error patterns defined through error codes, message keywords, and associated repair actions. Once an error is mapped to \texttt{RBR}, \toolname\ invokes the corresponding repair operator to transform the faulty SQL block before re-execution.

The key idea of RBR is to correct recurring failures through lightweight rewriting, without invoking full semantic reasoning. Depending on the matched error type, the repair may insert auxiliary statements, adjust the local statement context, or attach a recommended SQL template for subsequent repair stages. For example, when MySQL reports an error indicating that a table \texttt{was not locked with LOCK TABLES}, \toolname\ invokes an operator that prepends \texttt{UNLOCK TABLES;} to the faulty SQL block. Likewise, when MariaDB reports that a referenced table \texttt{does not exist}, \toolname\ can invoke a \texttt{createtable} operator that inserts a recommended \texttt{CREATE TABLE} template near the failing statement, thereby supplying structured context for later semantic repair.

Compared with directly invoking the LLM on every failing input, RBR provides a lightweight intermediate layer for high-frequency, structurally regular errors. Its role is not limited to fully repairing the input in one step; in many cases, it also prepares a more informative and constrained context for the subsequent intelligent semantic-aware repair stage. This design allows \toolname\ to recover quickly from common failures while reserving LLM-based reasoning for cases that genuinely require deeper semantic inference.

\subsubsection{Intelligent Semantic-Aware Repair}
After syntax- and error-aware filtering and rule-based repair, \toolname\ forwards the remaining repairable failures to an intelligent semantic-aware repair stage powered by an LLM. This stage handles two categories of cases: (1) inputs that remain unresolved after earlier lightweight repair, and (2) failures that are not filtered out as low-value errors but do not match any available rule-based repair pattern. Instead of discarding such inputs prematurely, \toolname\ attempts to recover them through context-sensitive semantic repair, so that structurally rich and potentially high-value test cases can continue to participate in fuzzing.

The rationale for this stage is that many DBMS failures, especially those involving under-tested features, cannot be corrected through local rewriting alone. Such failures often arise from incomplete semantic dependencies, including missing objects, unmet feature preconditions, incompatible session state, or invalid cross-statement ordering. These dependencies are difficult to enumerate exhaustively as hand-written rules, particularly across heterogeneous DBMS dialects and evolving feature sets. Therefore, \toolname\ employs an LLM as a practical mechanism for inferring missing execution context and restoring semantically valid feature invocations from the failing SQL fragment and DBMS diagnostic message.

For each unresolved case, \toolname\ constructs a structured repair prompt from the current SQL block, relevant schema or execution context, and the associated DBMS error message. The model is then asked to revise the test case so that the target feature or behavior can be exercised under a semantically valid context. Depending on the failure mode, the generated patch may introduce missing definitions, restore prerequisite statements, or rewrite incompatible clauses while preserving the original testing intent as much as possible. The repaired result is then re-injected into the fuzzing loop for re-execution.

Importantly, this repair process is inherently iterative. A single repair attempt may only partially resolve the original issue, and subsequent execution or mutation may further change the semantic context in which the test case is evaluated. As a result, \toolname\ does not treat semantic repair as a one-shot correction step, but as a feedback-driven refinement loop in which newly observed DBMS diagnostics continuously inform the next repair decision. Conceptually, this design is related to chain-of-thought-style decomposition: instead of requiring the model to recover the entire valid execution context in a single step, \toolname\ incrementally guides repair through successive error signals and evolving execution history. Notably, this iterative behavior arises at two levels: within the repair stage itself, where one repair may expose another unresolved dependency, and across the broader fuzzing loop, where subsequent mutations may alter the execution context and create new semantic requirements.

This stage is designed not merely to improve input validity, but to preserve difficult test cases that are more likely to exercise deeper control paths, feature interactions, and state-sensitive behaviors. In this sense, intelligent semantic-aware repair serves as the final recovery layer of \toolname\: it prevents semantically incomplete but high-value inputs from being discarded too early, thereby substantially improving the exploration of under-tested DBMS features.

\subsection{Replay-Guided Crash Validation}
\label{sec:Validation}
To ensure that detected crashes are genuinely reproducible and to determine whether they depend on persistent DBMS state accumulated across executions, \toolname\ implements a replay-guided crash validation pipeline followed by test case reduction. The goal of this stage is not only to confirm the existence of a crash, but also to identify whether the failure emerges from cross-statement or cross-execution state interactions and to extract a minimal yet effective proof-of-concept (PoC).

\textbf{Replay-based crash validation.} 
\toolname\ deliberately preserves session-level execution history during fuzzing because many DBMS failures, especially those involving under-tested features, are not localized to a single standalone test case. Instead, they often emerge only after prior statements have modified internal metadata, runtime modes, replication state, or other persistent execution context. To support this behavior, \toolname\ resets the logical database namespace using \texttt{DROP DATABASE} and \texttt{CREATE DATABASE}, while intentionally retaining the broader session context across executions. In this sense, replay-guided validation is not merely a debugging aid, but part of the bug-finding model of \toolname.

When a crash is detected, \toolname\ first re-executes the crashing test case in isolation to determine whether the failure is self-contained. If the crash does not reproduce, \toolname\ replays the recorded execution history by re-executing prior statements in their original order before the crashing input. If the failure only manifests under such replay, the bug is classified as state-dependent. Rather than passively relying on incidental residual effects, \toolname\ actively records crashing inputs together with their preceding execution history and validates them under controlled replay prefixes, so that failures rooted in persistent state interactions can be distinguished from purely local crashes.

\textbf{Dependency-preserving Test Case Reduction.}
Once a crash is confirmed, \toolname\ automatically performs test case reduction to produce a minimal proof-of-concept (PoC). This process targets two dimensions: (1) structural simplification of individual SQL statements, and (2) reduction of the overall statement sequence. These two simplification strategies are applied iteratively and alternately, progressively minimizing the test case while ensuring the crash remains reproducible.

Reduction begins from the last statement in the sequence—where the crash is typically triggered—and proceeds backward. For structural simplification, each SQL statement is first parsed into an abstract syntax tree (AST), and then recursively simplified by pruning subtrees or nodes. The simplified AST is converted back into SQL and tested. If the crash still occurs, the simplification is retained. Otherwise, the system rolls back the change and attempts alternative simplifications.
Once structural simplification weakens or removes certain data or control dependencies, the surrounding context becomes amenable to further reduction. At this stage, \toolname\ attempts to delete preceding statements using a delta debugging strategy, preserving only those necessary for triggering the fault. The final output is a compact, semantically valid SQL sequence that isolates the root cause of the crash.

\toolname’s replay and reduction workflow reconstructs the full crash path and eliminates irrelevant logic, producing a concise PoC of critical SQL statements, thereby improving debugging clarity and the quality of bug reports.

\section{Implemention}
\subsection{Prompt Engineering}
To ensure stable generation quality and precise control over the output, \toolname\ uses structured prompt templates tailored to SQL statement instantiation and semantic error repair. 
The complete prompt templates used in our implementation are provided in Appendix~\ref{appendix:prompt}.

\subsubsection{Prompt Structure for SQL Instantiation}
The goal of SQL instantiation is to transform grammar-expanded abstract templates into concrete, executable SQL queries. \toolname\ constructs prompts with the following four key components:
\begin{itemize}[leftmargin=1.8em]
	\item \textbf{Task Instruction}: A concise natural language directive that instructs the LLM to generate a syntactically valid and semantically meaningful SQL query.
	
	\item \textbf{Schema Initialization Statements}: Randomly generated \texttt{CREATE TABLE/VIEW} statements that define the database context. These provide schema and data type grounding (e.g., table names, column types, constraints) for coherent query generation.
	
	\item \textbf{SQL Templates}: A set of SQL templates derived from the grammar, serving as a structural constraint.

	\item \textbf{Output Formatting Instruction}: A final constraint that directs the LLM to return only complete SQL statements, avoiding comments, explanations, or multiple alternatives.
\end{itemize}
This structured prompting allows the LLM to fill in semantically appropriate content, such as meaningful column selections, valid \texttt{WHERE} clauses, or realistic \texttt{JOIN} clauses, all consistent with the provided schema.

\subsubsection{Prompt Structure for Error Repair}
If a SQL test case fails with errors addressable by semantic-aware repair, \toolname\ constructs an LLM repair-instruction prompt containing:

\begin{itemize}[leftmargin=1.8em]
	\item \textbf{Repair Instruction}: A natural language request that instructs the model to identify the faulty SQL segment and either modify it or insert the required definitions to ensure successful execution.

	\item \textbf{Full SQL Context}: The complete test case, including schema setup and execution sequence, providing the necessary context for understanding the failure and synthesizing a valid patch.
	
	\item \textbf{Error Annotations}: The error message returned by the DBMS is injected near the offending SQL line, using inline comments.
	
	\item \textbf{Output Formatting Instruction}: Follow the same formatting rules as those applied during the \textit{Instantiation} phase.
\end{itemize}

The prompt structure allows the LLM to reason about schema-object relationships, identify the cause of failure, and propose actionable fixes, such as defining a missing object or correcting column usage, without discarding the broader intent of the test case.

\subsection{Execution Framework and Instrumentation}
\toolname\ supports fuzzing for both client-server and embedded DBMS architectures through adaptive execution strategies tailored to the target under test. Instead of conventional stdin-based fuzzing, it delivers SQL inputs via realistic interaction paths to preserve semantic context and execution fidelity.

\subsubsection{DBMS Fuzzing Drivers}
\begin{sloppypar}
	\toolname\ supports both client-server and embedded DBMS architectures through tailored execution strategies.
	For client-server systems (e.g., MySQL, MariaDB), \toolname\ launches the command-line interface client via \texttt{pexpect.spawn()} and maintains a persistent session. To isolate test cases without restarting the server, it resets the environment using a standard SQL prelude: \texttt{DROP DATABASE IF EXISTS test\_db; CREATE DATABASE test\_db; USE test\_db;}.
	This ensures a clean database context while preserving session-level state, enabling discovery of state-dependent bugs beyond fork-per-input fuzzers.
	For embedded engines (e.g., SQLite), \toolname\ uses a forkserver-style subprocess to run the SQLite shell. It alternates between temporary and active database files (via .open tmp.db and .open test.db) to simulate a clean environment with minimal overhead, enabling efficient batch fuzzing without shell restarts.
\end{sloppypar}

\subsubsection{Instrumentation and Coverage Feedback}
All DBMS binaries are compiled with AFL++'s afl-clang-fast or afl-clang-fast++ toolchain to enable edge coverage instrumentation. AddressSanitizer (ASan) is also enabled to capture memory-related faults such as buffer overflows or use-after-free errors. No manual modifications to DBMS source code are required, allowing compatibility with standard upstream builds.
Coverage feedback from AFL++ guides the mutation engine by prioritizing test cases that explore novel code paths. Only inputs that contribute new coverage are retained, while others are discarded.

\subsubsection{Fuzzing Strategy and Determinism}
\toolname\ executes test cases in a strictly single-threaded fashion to maintain determinism, which is essential for reliable crash replay and debugging.
Unlike AFL-style forkserver fuzzing, \toolname\ maintains a persistent DBMS client and executes test inputs sequentially.
This design ensures that the entire execution history is preserved and can be replayed when a crash occurs.
In contrast, multi-threading is employed only for interacting with the LLM backend, where multiple concurrent threads handle prompt generation, SQL instantiation, and error repair. This architectural separation allows \toolname\ to balance deterministic fuzzing with high-throughput generation.
Crash detection relies on monitoring signal-based terminations (e.g., SIGSEGV, SIGABRT), abnormal exit codes, and standard error logs. Inputs that trigger these crashes, along with those that increase coverage, are retained in the seed pool for further mutation and replay.
By combining persistent-mode execution, lightweight state resets, and compiler-based instrumentation, \toolname\ achieves efficient and scalable fuzzing across diverse DBMS environments.

\section{Evaluation}

\subsection{Experimental Setup}
All experiments are carried out on a server equipped with an Intel Xeon Gold 6430 CPU (32 cores), 256 GB RAM, and a NVIDIA A100 GPU, running Ubuntu 20.04 LTS. The GPU resources are reserved exclusively for local inference of LLMs. 
To ensure broad applicability, we evaluate \toolname\ on five representative open source database systems that collectively cover both client-server and embedded DBMS architectures. Specifically, we target MySQL 9.1.0, MariaDB 11.4.0, SQLite 3.47.2, PostgreSQL 17.2, and ClickHouse v24.12.2.29-stable.
\toolname\ integrates the LLM model via a locally deployed instance of Qwen3-30B, running on the vLLM inference framework. 
During vulnerability mining, we also tested several advanced models; however, for controlled and reproducible evaluation, all reported results are based on Qwen3-30B.
We used a temperature of 0.3–0.5 to balance stability and diversity. In pilot tests, a temperature of 0.4 produced natural yet varied outputs—sufficiently diverse for fuzzing under structured constraints. All experiments used an 8K-token context window.

To demonstrate the advantage of \toolname, we compare it against three state-of-the-art open-source fuzzing baselines: Squirrel~\cite{27zhong2020squirrel}, EET~\cite{18jiang2024detecting}, and SQLancer~\cite{11rigger2020detecting}. Among them, Squirrel represents mutation-based fuzzing, while both EET and SQLancer adopt generation-based approaches. 
Since EET and SQLancer are not inherently a grey-box fuzzing tool, we first collect their generated test cases and then utilize \toolname's replay mechanism to measure its coverage. For SQLancer, we use TLP~\cite{12rigger2020findingbugs} as the test oracle for MySQL, and NoREC~\cite{11rigger2020detecting} for the other DBMS targets. This choice is constrained by the design of SQLancer, which employs different oracle-based testing strategies tailored to specific DBMSs.
All baselines were evaluated under identical runtime conditions and DBMS versions.
We adopted official configurations and standardized crash triage procedures to ensure fair and reproducible comparisons.
Except for Squirrel, a mutation-based fuzzer using its default seeds, all baselines ran without a seed corpus.
Each experiment was repeated 5 times on identical hardware, and the reported results are the averages, which show consistent statistical stability.

\begin{table}[tbp]
	\setlength{\abovecaptionskip}{4pt}
	\setlength{\tabcolsep}{4pt}
	\renewcommand{\arraystretch}{0.94}
	\centering
	\small
	\caption{Bugs discovered by \toolname}
	\label{tab:bugs}
	
	\begin{tabular*}{\columnwidth}{@{\extracolsep{\fill}}ccccc}
		\toprule
		\textbf{DBMS} & \textbf{Found} & \textbf{Confirmed} & \textbf{Fixed} & \textbf{Feature-Related} \\
		\midrule
		MySQL       & 22 & 22 & 12 & 10 \\
		MariaDB     & 28 & 24 & 10  & 9 \\
		SQLite      & 2  & 2  & 2  & 1 \\
		ClickHouse  & 12  & 12  & 7  & 7 \\
		\midrule
		\textbf{Total} & \textbf{64} & \textbf{60} & \textbf{31} & \textbf{27} \\
		\bottomrule
	\end{tabular*}
	
\end{table}

\subsection{Bug Discovery Effectiveness}
To evaluate the practical effectiveness of \toolname, we conducted a systematic fuzzing campaign on five representative open-source DBMSs. In total, \toolname\ discovered 64 distinct bugs, among which 60 have been confirmed by vendors and 9 have been assigned CVE identifiers. As summarized in Table~\ref{tab:bugs}, 31 issues have already been fixed, while the remaining ones are scheduled for upcoming releases. Notably, 27 of the confirmed bugs are related to under-tested DBMS special features. This observation suggests that \toolname\ does not merely improve generic SQL fuzzing coverage, but is particularly effective at exposing failures rooted in feature-specific and under-explored DBMS behaviors.

\begin{figure}[htbp]
	\setlength{\abovecaptionskip}{1.5pt}
	\centering
	\includegraphics[width=0.98\columnwidth]{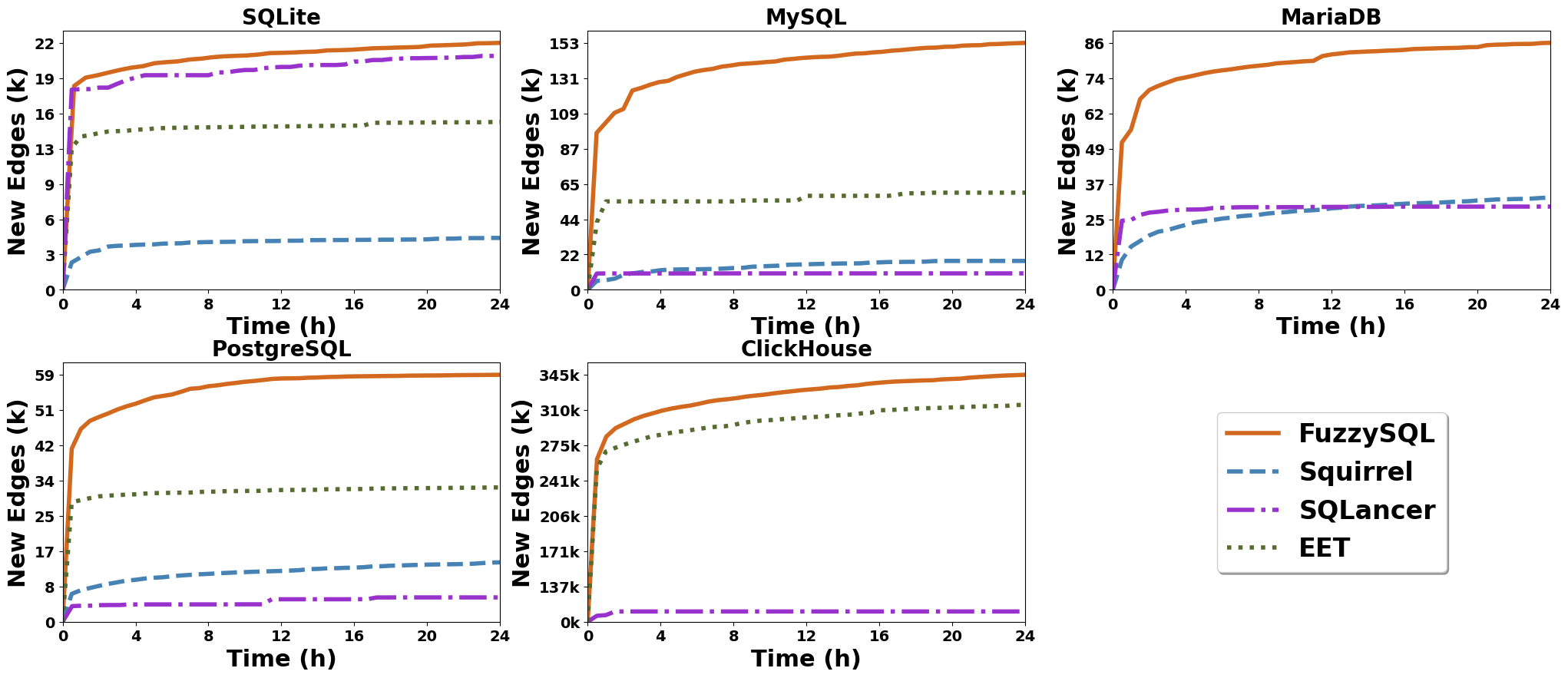}
	\caption{The number of new edges discovered in 24 hours.}
	\label{fig:coverage}
	
\end{figure}

Beyond long-term bug discovery, we further compare fuzzing effectiveness under a fixed 24-hour budget. As shown in Figure~\ref{fig:coverage}, \toolname\ consistently outperforms all baselines across the evaluated DBMSs in terms of newly discovered execution edges. The advantage is especially pronounced on MySQL, MariaDB, and PostgreSQL, where \toolname\ achieves substantially higher coverage than the strongest baseline. On SQLite and ClickHouse, the margin is smaller. This is likely because SQLite exposes a relatively lightweight feature space, while ClickHouse already benefits from EET's generation of deeply nested analytical queries. Even so, \toolname\ still attains the highest overall coverage across all five systems.

To understand how this coverage advantage translates into concrete vulnerability discovery, Table~\ref{tab:mysql-24bugs} reports deduplicated MySQL bugs discovered within 24 hours and shows which fuzzers were able to trigger each bug. Among the evaluated baselines, only EET uncovers a limited number of bugs, while none of the baselines expose feature-related vulnerabilities in this setting. In contrast, \toolname\ consistently triggers both feature-related and non-feature-related bugs under the same time budget. This result indicates an important capability gap between \toolname\ and prior DBMS fuzzers: although existing approaches can occasionally detect generic crashes, they are considerably less effective at constructing and evolving the semantically valid, state-sensitive SQL sequences required to exercise under-tested DBMS functionality.

Overall, these results show that the effectiveness of \toolname\ lies not only in finding more bugs or reaching higher coverage, but also in exposing a broader and qualitatively different set of vulnerabilities. In particular, feature-related bugs are not isolated anecdotes in our results; with 27 confirmed cases, they constitute a substantial portion of the bugs uncovered by \toolname. This capability arises from \toolname's integrated design, which jointly addresses three obstacles in feature-related DBMS bug discovery: insufficient coverage of feature-specific constructs, unmet semantic preconditions, and hidden state interactions that span multiple statements or executions.

\begin{table}[htbp]
	\setlength{\abovecaptionskip}{4pt}
	\centering
	\small
	
	\begin{threeparttable}
		\caption{Deduplicated MySQL bugs discovered within 24 hours}
		\label{tab:mysql-24bugs}
		
		\providecommand{\cmark}{\smash{\scriptsize\textcolor{green!60!black}{\ding{52}}}}
		\providecommand{\xmark}{\smash{\scriptsize\textcolor{red}{\ding{55}}}}
		
		\renewcommand{\arraystretch}{0.94}
		\setlength{\tabcolsep}{3pt}
		
		\begin{tabular*}{\columnwidth}{@{\extracolsep{\fill}}cccccccc|cc}
			\toprule
			\textbf{Target} & \textbf{ID$\dagger$} & \textbf{FR} & \textbf{Type} &
			\rotatebox{90}{\textbf{FuzzySQL}} &
			\rotatebox{90}{\textbf{Squirrel}} &
			\rotatebox{90}{\textbf{EET}} &
			\rotatebox{90}{\textbf{SQLancer}} &
			\rotatebox{90}{\textbf{FuzzySQL$^{\text{!r}}$}} &
			\rotatebox{90}{\textbf{FuzzySQL$^{\text{!r}}_{\text{!m}}$}} \\
			\midrule
			MySQL & 3  & \frNo  & Use After Free    & \cmark & \xmark & \cmark & \xmark & \xmark & \xmark \\
			MySQL & 6  & \frNo  & Assertion Failure & \cmark & \xmark & \xmark & \xmark & \cmark & \xmark \\
			MySQL & 8  & \frYes & Assertion Failure & \cmark & \xmark & \xmark & \xmark & \cmark & \xmark \\
			MySQL & 9  & \frNo  & NULL Ptr Deref    & \cmark & \xmark & \cmark & \xmark & \cmark & \cmark \\
			MySQL & 17 & \frYes & Assertion Failure & \cmark & \xmark & \xmark & \xmark & \xmark & \xmark \\
			MySQL & 21 & \frNo  & NULL Ptr Deref    & \cmark & \xmark & \cmark & \xmark & \cmark & \xmark \\
			\bottomrule
		\end{tabular*}
		
		\begin{tablenotes}[flushleft]
			\footnotesize
			\item $\dagger$ Bug ID corresponds to Table~\ref{tab:vuln_status} (Appendix~\ref{appendix:vuln_status}). FR: \frYes\ feature-related, \frNo\ not feature-related; \textbf{!r}: repair disabled; \textbf{!m}: mutation disabled. 
		\end{tablenotes}

	\end{threeparttable}
\end{table}

\subsection{Ablation Study}
We perform an ablation study to assess the impact of three core design choices in \toolname: logic-shifting progressive mutation, automated semantic repair, and the choice of LLM for semantic instantiation. Our analysis addresses the following questions:

\begin{itemize}[leftmargin=1.8em]
	\item How critical is the logic mutation for triggering complex bugs?
	\item Does the error repair pipeline meaningfully improve fuzzing effectiveness despite its computational cost?
	\item How significantly does the choice of LLM affect the quality of generated test cases?
\end{itemize}

\begin{figure}[htbp]
	\setlength{\abovecaptionskip}{2pt} 
	\centering
	\includegraphics[width=0.65\textwidth]{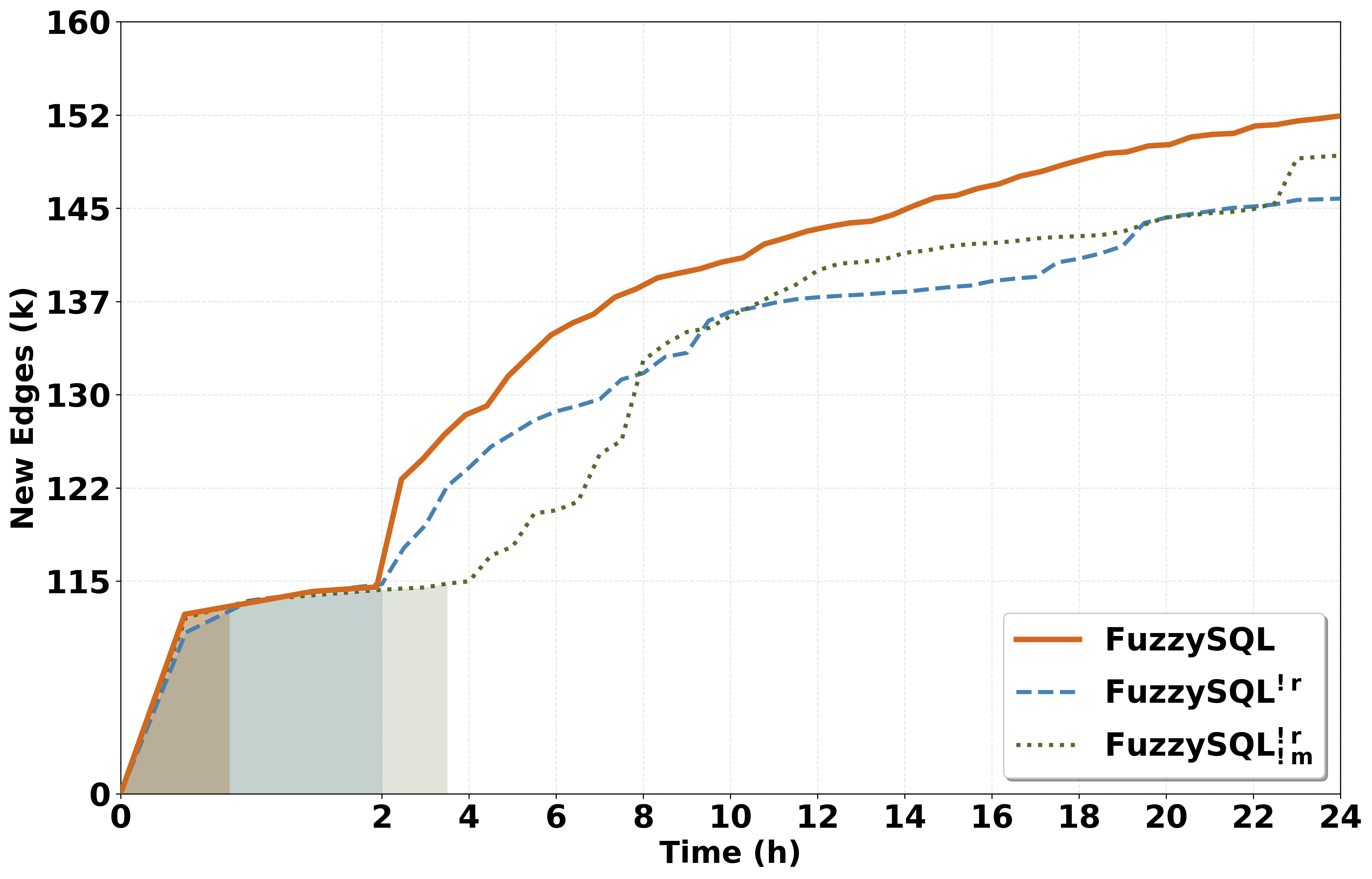}
	\caption{Ablation Study of FuzzySQL Components.}
	\label{fig:ablation}

\end{figure}

\textbf{System-Level Impact of Core Components.}
Figure~\ref{fig:ablation} illustrates the system-level impact of logic-shifting mutation and semantic repair on both coverage growth and early bug discovery. All configurations exhibit comparable coverage gains during the initial $\sim$2 hours, indicating that grammar-guided generation is sufficient to bootstrap shallow execution paths. However, clear divergence emerges as fuzzing progresses: the full \toolname\ consistently advances faster and achieves the highest final coverage (152.7k edges), compared to 146.0k for \toolname$^{\text{!r}}$ and 149.5k for \toolname$^{\text{!r}}_{\text{!m}}$. This gap suggests that generation alone is insufficient for sustained exploration of deeper and more stateful DBMS behaviors.

The benefit of these components is more pronounced when considering bug discovery efficiency. As shown by the shaded regions, \toolname\ triggers its first confirmed bug within approximately 0.83 hours, whereas disabling semantic repair delays the first bug to 2 hours, and disabling both repair and mutation further postpones discovery to 3.5 hours. Although \toolname$^{\text{!r}}$ and \toolname$^{\text{!r}}_{\text{!m}}$ reach similar final coverage, logic-shifting mutation clearly improves early-phase exploration and accelerates vulnerability discovery, rather than merely increasing eventual coverage.

\begin{sloppypar}
	This observation aligns with Table~\ref{tab:mysql-24bugs}. Under the same 24-hour budget, only the full \toolname\ reliably discovers feature-related bugs and uncovers more unique vulnerabilities overall, while ablated variants miss several feature-dependent failures or exhibit reduced bug-finding capability. We attribute this difference to \toolname’s integrated pipeline design, in which grammar-guided generation, progressive logic-shifting mutation, and semantic-aware repair jointly construct, evolve, and recover diverse execution contexts. Together, these components enable systematic exploration of feature-specific and state-dependent control paths and convert them into reproducible vulnerabilities, beyond what any individual component can achieve in isolation.
\end{sloppypar}

\begin{figure}[htbp]
	\setlength{\abovecaptionskip}{2pt} 
	\centering
	\includegraphics[width=0.65\textwidth]{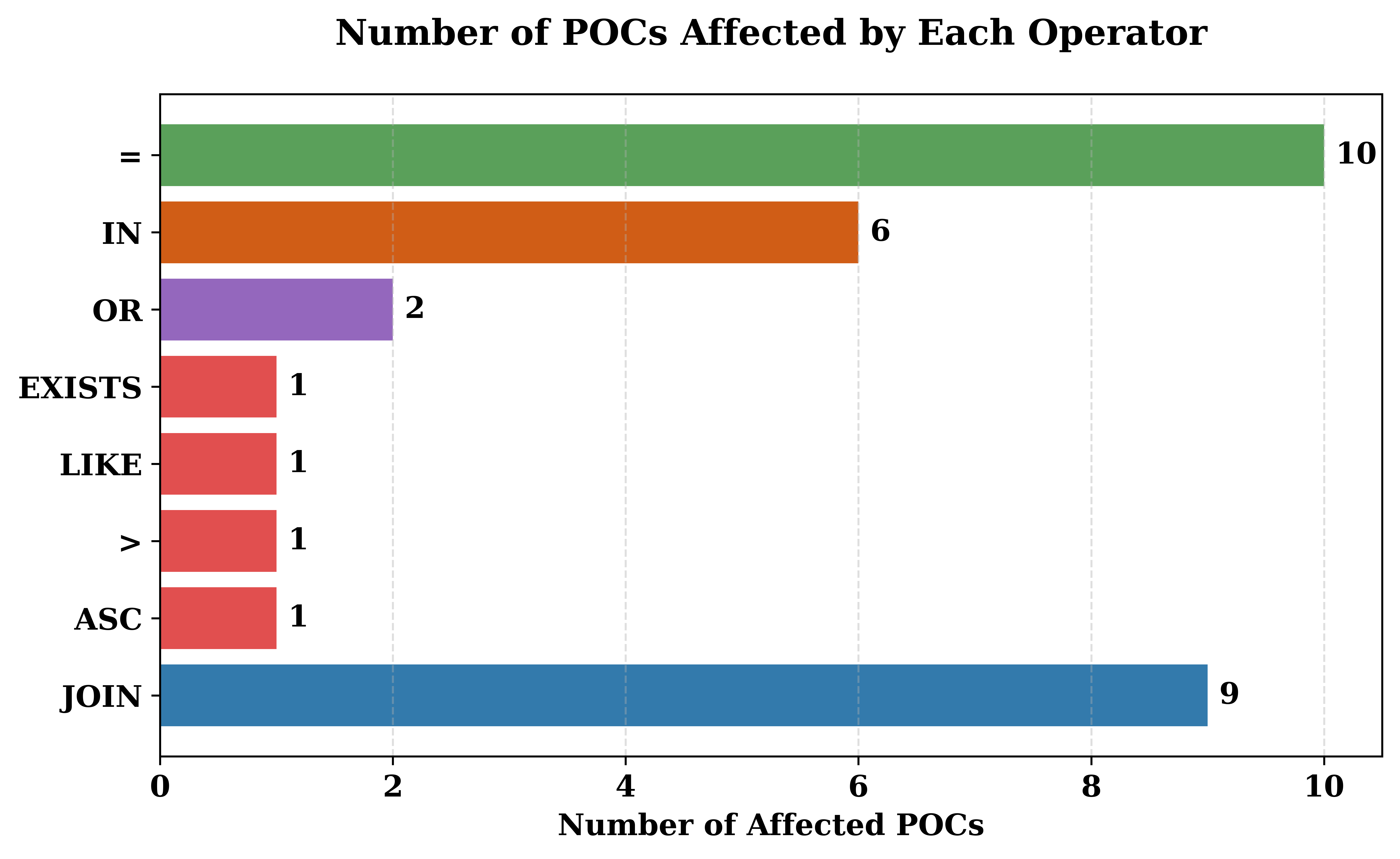}
	\caption{Impact of Logic-shifting on POC Effectiveness.}
	\label{fig:logic_shift_types}

\end{figure}

\textbf{Effectiveness of Logic Mutation.}
We analyze all 60 confirmed bugs discovered by \toolname\ to identify those that depend on logic-shifting transformations. A bug is considered logic-sensitive if its PoC contains a condition (e.g., \texttt{=}, \texttt{IN}, \texttt{JOIN}) mutated from another semantically valid form and becomes non-triggering when reverted. Using this criterion, 22 out of 60 bugs (36.67\%) are logic-sensitive, indicating that logic mutation is a major contributor to vulnerability discovery.
Figure~\ref{fig:logic_shift_types} summarizes the distribution of these bugs by transformation type. Equality polarity changes  account for the largest fraction (10 PoCs), followed by \texttt{JOIN}-related rewrites (9 PoCs) and membership predicates \texttt{IN}-related  rewrites (6 PoCs). Other transformations, including \texttt{AND}/\texttt{OR} rewrites, ordering reversals, and \texttt{EXISTS} manipulation, contribute additional unique triggers. Together, these results show that small, semantics-preserving logic shifts can effectively expose alternative execution paths that are rarely exercised by structure-preserving or generation-only fuzzers.

\textbf{Effectiveness of Automated Repair.}
Although semantic-aware repair introduces additional overhead, it substantially improves fuzzing yield. In our evaluation, 16 out of the discovered bugs originate from test cases that were initially invalid and would have been discarded without repair, including cases with missing objects or unmet semantic preconditions.
Consistent with the system-level results and Table~\ref{tab:mysql-24bugs}, disabling repair reduces the number of unique bugs discovered and causes several feature-related failures to be missed. These results indicate that automated repair is essential for preserving and exploiting high-value test cases, rather than serving as a mere convenience for handling invalid inputs.

\begin{figure}[htbp]
	\setlength{\abovecaptionskip}{2pt} 
	\centering
	\includegraphics[width=0.8\textwidth]{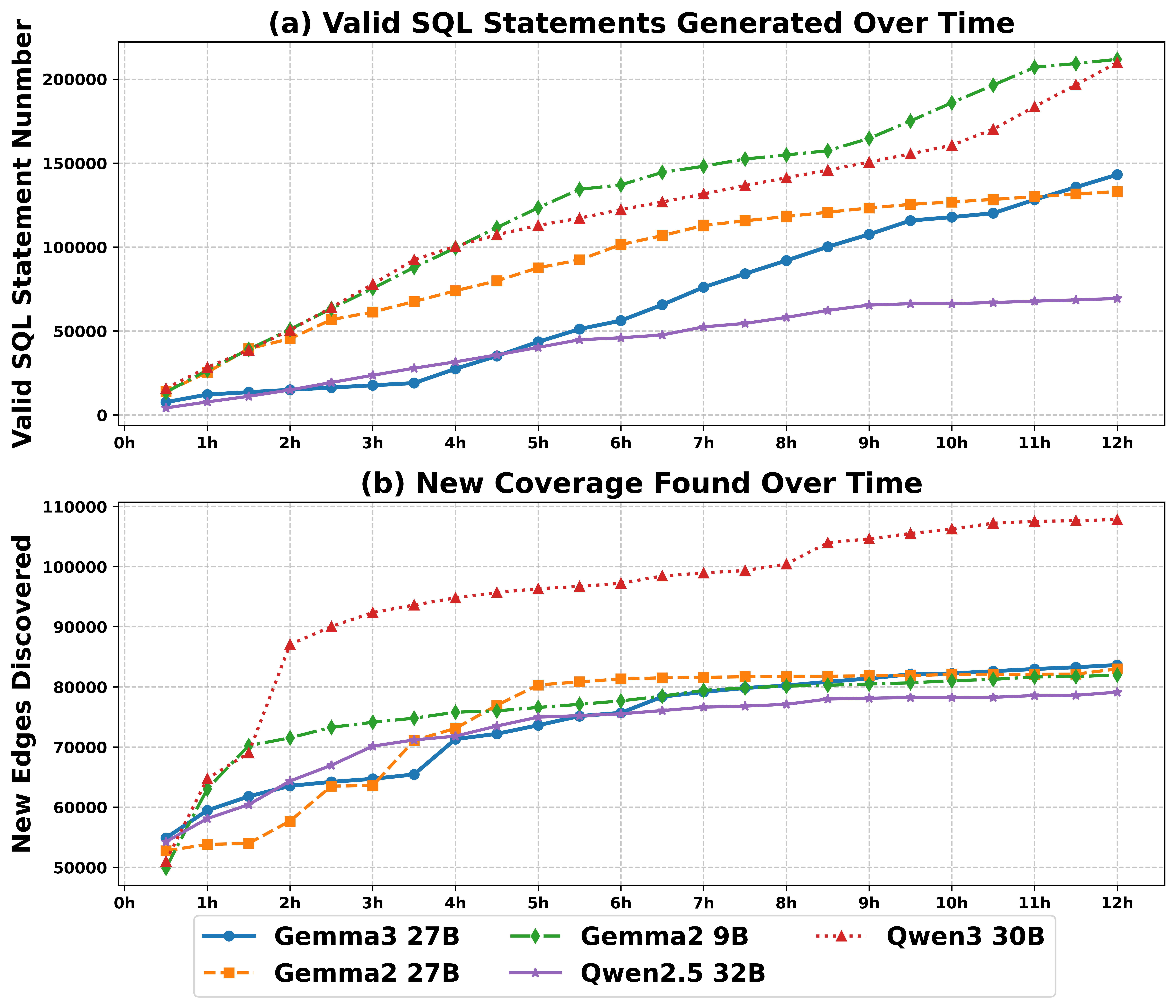}
	\caption{Effectiveness of Different LLMs.}
	\label{fig:llmchoice}
\end{figure}

\textbf{Impact of LLM Choice.}
We conducted an ablation study on MySQL to assess the impact of different LLMs on \toolname’s performance. We evaluated Gemma3-27B, Gemma2-27B, Gemma2-9B, Qwen2.5-32B, and Qwen3-30B, 
each running on an A100 GPU with a context length of 8192. Over a 12-hour period, we recorded the number of valid SQL statements generated (Figure \ref{fig:llmchoice}a) and the number of new execution edges discovered (Figure \ref{fig:llmchoice}b).
Figure \ref{fig:llmchoice}a shows that Gemma2-9B produces the largest number of valid SQL statements with the highest throughput and efficiency. However, Figure \ref{fig:llmchoice}b reveals that this advantage does not translate into deeper code exploration. 
In contrast, Qwen3-30B generates fewer statements overall but consistently achieves the highest coverage growth over time. This indicates that its outputs are of higher semantic quality, capable of exercising deeper and more diverse program behaviors.

These results demonstrate that stronger models, such as Qwen3-30B, produce higher-quality and more diverse test cases, leading to greater coverage and bug-discovery potential.
However, we opt for locally deployed models rather than API-based SoTA ones (e.g., GPT-4o) due to cost and scalability concerns—running such models at fuzzing scale would incur thousands of dollars and hundreds of compute hours.
Similarly, reasoning-oriented models (e.g., OpenAI-o1, DeepSeek-R1) were excluded because their multi-step reasoning severely limits throughput, making them unsuitable for large-scale fuzzing.
Therefore, \toolname’s design favors efficient, instruction-aligned local LLMs, which strike a balance between semantic quality and generation efficiency for large-scale fuzzing workloads.

\subsection{Case Study}
\label{subsec:Casestudy}
To complement our quantitative evaluation, we present case studies that demonstrate the effectiveness of \toolname\ in three distinct dimensions. These examples illustrate how the system exposes critical vulnerabilities in the DBMS internals by leveraging feature-specific generation, semantic-aware mutation, and error recovery. Each case highlights a different challenge in DBMS fuzzing and how \toolname\ overcomes it through its design.

\textbf{Feature-related bugs in under-tested DBMS functionalities.} We present two representative vulnerability-triggering inputs found in MySQL and MariaDB,
which are shown in Listing \ref{lst:mysql-analyze}, Listing \ref{lst:mariadb-kill} and \ref{lst:mariadb-proc} respectively.
These cases violate the standard SQL syntax or DBMS usage rules—MySQL does not support the combined use of \texttt{DESC} and \texttt{FORMAT}; MariaDB's \texttt{KILL} statement lacks support for predicate logic; and the incorporation of composite subquery expressions within \texttt{PROCEDURE} declarations in MariaDB is improper. As a result, traditional DBMS fuzzers, often guided by hand-crafted grammars and strict semantic correctness, are unlikely to generate such inputs.

In contrast, \toolname\ leverages grammar-driven structural expansion combined with LLM-based forced instantiation to generate atypical, boundary-pushing SQL inputs. By decoupling structure from content, these "fuzzy templates" serve as scaffolds that guide the LLM beyond its default bias toward well-formed, canonical queries—a limitation inherited from pretraining on normative data. This mechanism encourages the generation of rule-bending or underspecified statements that lie outside conventional syntax but remain executable.
Such flexibility cannot be easily achieved through manually crafted grammars. For example, the combination of \texttt{KILL} with predicate logic in Listing~\ref{lst:mariadb-kill} represents a usage pattern that is both rare and counterintuitive, something human-authored templates are unlikely to anticipate, and traditional generators lack the means to instantiate. \toolname's design, by contrast, allows it to discover these error-prone interactions naturally through guided structural diversity and semantic inference.

\vspace{0.2cm}

\begin{minipage}{0.97\textwidth}
	\begin{lstlisting}[style=sqlstyle, caption={MySQL crash via invalid DESC ANALYZE syntax.}, label=lst:mysql-analyze]
		CREATE TABLE t1 (v1 JSON, v2 CHAR(7), v3 BINARY(3));
		DESC ANALYZE FORMAT = JSON (SELECT v1 FROM t1 UNION XXXX);
		
	\end{lstlisting}
	
\end{minipage}

\begin{minipage}{0.97\textwidth}
	\begin{lstlisting}[style=sqlstyle, caption={MariaDB crash via malformed KILL expression.}, label=lst:mariadb-kill]
		KILL 1 IN (SELECT 1) IN (XXXX);
	\end{lstlisting}
	
\end{minipage}

\begin{minipage}{0.97\textwidth}
	\begin{lstlisting}[style=sqlstyle, caption={MariaDB crash via subquery misuse in PROCEDURE.}, label=lst:mariadb-proc]
		CREATE PROCEDURE test_proc (id VARCHAR(255)) BEGIN DECLARE dt
		DATETIME(6) DEFAULT ROW(XXXX) = SOME(SELECT 1) = ALL(XXXX);
	\end{lstlisting}
	
\end{minipage}

\begin{minipage}{0.97\textwidth}
	\begin{lstlisting}[style=sqlstyle, caption={MySQL crash via malformed INSTALL COMPONENT inducing conditional logic.}, label=lst:logic-sensitive]
		INSTALL COMPONENT 'plugin2' SET GLOBAL default.variable1 = ON, t0.c1 = 1 OR 1 = ANY (XXXX);
	\end{lstlisting}
	
\end{minipage}

\begin{minipage}{0.97\textwidth}
	\begin{lstlisting}[style=sqlstyle, caption={Repaired SQLite input that originally failed due to ambiguous column references.}, label=lst:sqlite-repair]
		CREATE TABLE t1 (XXXX);
		CREATE TABLE t2 (XXXX) WITHOUT ROWID;
		INSERT INTO t2 (XXXX) VALUES (XXXX);
		ANALYZE t2;
		WITH cte_0 AS (SELECT 1) SELECT DISTINCT ref_2.v1 FROM (cte_0 JOIN generate_series(79, 105, 7) AS ref_1 ON (ref_1.value = 1)) CROSS JOIN (t2 AS t2a JOIN (t1 AS ref_2 JOIN t2 ON (ref_2.v2 = 1))); -- Error: ambiguous column name: main.t2.v1
		WITH cte_0 AS (SELECT 1) SELECT DISTINCT ref_2.v1 FROM (cte_0 JOIN generate_series(79, 105, 7) AS ref_1 ON (ref_1.value = 1)) CROSS JOIN (t2 AS t2a JOIN (t1 AS ref_2 JOIN t2 as ref_3 ON (ref_2.v2 = 1))); -- Fixed and Trigger Crash
	\end{lstlisting}
	
\end{minipage}

\begin{sloppypar}
	\textbf{Logic-sensitive bugs enabled by conditional mutation.}
	\toolname’s lightweight logic mutation enables subtle yet effective transformations in SQL conditions, allowing it to explore alternative control paths often missed by traditional fuzzers. A representative example is shown in Listing~\ref{lst:logic-sensitive}, where a conditional expression inside an \texttt{INSTALL COMPONENT} statement includes a comparison of the disjunctive subquery. This particular crash is triggered only  when the clause \texttt{t0.c1 = 1} is present—replacing it with \texttt{t0.c1 != 1} or any other variation prevents the assertion failure (as confirmed by GDB stack trace analysis).
	This demonstrates that the bug is not merely exposed through structural coverage, but depends on a precise logical condition being satisfied during execution. Existing mutation strategies, whether focused on AST structures or statement-level heuristics, often overlook such fine-grained control flow variations and struggle to achieve this level of logical precision. By systematically shifting operators, \toolname\ introduces semantic perturbations that preserve syntax while altering the evaluation paths. This approach not only increases test diversity, but also surfaces fragile logic handling within DBMS internals.

\end{sloppypar}

\textbf{Repaired bugs salvaged from initially invalid test cases.}
\toolname’s semantic-aware repair module enables recovery of otherwise invalid inputs that can still uncover real vulnerabilities. Listing~\ref{lst:sqlite-repair} shows a PoC that triggers a heap-buffer-overflow in SQLite during join resolution in \texttt{sqlite3WhereEnd()}. Initially, this input failed with the error “\texttt{ambiguous column name: main.t2.v1}” due to multiple unaliased uses of table \texttt{t2}, which caused semantic ambiguity during join resolution.
Traditional fuzzers typically discard such malformed cases. Instead, \toolname\ follows the SAR Violate Constraints error handling strategy to capture error message via tagging and incorporates it into a structured LLM prompt. The model identifies the ambiguity and suggests a minimal fix—assigning an alias to one instance of \texttt{t2}—which enables execution and exposes the underlying bug.
This example highlights how \toolname’s repair pipeline can salvage high-value test cases that would otherwise be lost, turning parse-time failures into exploitable paths through lightweight, context-aware correction.

\section{Discussion}
This section discusses the broader implications, strengths, and limitations of \toolname. This work not only delivers concrete results in uncovering functionality-related vulnerabilities in DBMSs, but also offers new perspectives on how LLMs can reshape the design of fuzzing frameworks.

\textbf{Uncovering Hidden Vulnerabilities in DBMS Special Features.}
Our findings reveal that modern DBMSs expose a wide range of special-purpose SQL features—such as replication control, logging, and statistical commands—that are rarely targeted by traditional fuzzers. Although infrequent in everyday workloads, these features are tightly integrated into critical control paths and can cause severe runtime failures when misused or exercised under uncommon conditions.
\toolname\ systematically explores this under-tested space by generating multi-statement SQL sequence that simulate realistic yet edge-case feature combinations. This exposes control-path inconsistencies and metadata mismanagement bugs that are not triggered by typical DML/DDL queries. These results highlight a critical blind spot in conventional fuzzing techniques, which tend to focus on common syntax patterns and fail to explore the operational depth of feature-rich systems.

\begin{sloppypar}
	\textbf{Flexibly Adapt to Diverse DBMS Targets.}
	To migrate \toolname\ to a new DBMS, users only need to provide the target’s grammar specification (e.g., an ANTLR4 \texttt{.g4} file), from which \toolname\ automatically extracts syntax rules for template-based SQL generation. In addition, \toolname\ requires a lightweight database initialization component to generate basic \texttt{CREATE} and \texttt{INSERT} statements. This component involves enumerating supported data types of the target DBMS and applying largely uniform schema and data generation rules, which introduces minimal manual effort and is straightforward to implement.
	For error repair, users collect execution feedback during an initial fuzzing phase and map observed errors to high-level categories. Most repair strategies are DBMS-agnostic and can be reused across systems, requiring little to no customization for new targets. 
	As a result, supporting new DBMS targets requires limited manual customization.

\end{sloppypar}

\textbf{Revisiting the Role of LLMs in Fuzzing.}
A key insight of this work is that LLMs, when guided by structured grammar templates and diverse execution contexts, can serve as more than just language generators—they can be used as semantic reasoning engines.
Rather than relying solely on random AST-level mutations, \toolname\ uses LLMs to instantiate meaningful programs based on context, such as generating valid stored procedure bodies or logically connected statement sequences. This allows fuzzing to move beyond surface-level syntax exploration and into deeper logic modeling.
By decoupling structural constraints (enforced by templates) from content generation (handled by LLM), we retain structural diversity while still harnessing the semantic strengths of LLMs. 
This design helps mitigate LLMs' bias toward producing safe, canonical outputs and encourages exploration of more unusual and error-prone cases.

\textbf{Limitations and Future Outlook.}
While \toolname\ is effective in exposing semantic and control-flow bugs, it currently underperforms compared to EET in metrics such as statement depth and nesting complexity. 
This stems from inherent limitations in current LLMs, which tend to struggle with generating deeply nested or highly recursive SQL structures. Additionally, LLM-generated inputs occasionally require repair due to inconsistencies between declared objects and later usage, prompting our design of a repair pipeline.
However, these challenges largely reflect the limitations of current LLMs. As LLMs improve in reasoning depth and code synthesis capability, their effectiveness in fuzzing high-complexity structures will improve correspondingly.
\toolname\ excels at triggering feature-related failures and hard-to-reach logic branches, aspects that are often invisible to basic coverage models.

\section{Related Work}
\subsection{DBMS Fuzzing}

Fuzzing has emerged as a critical technique for testing database management systems (DBMSs) by directly targeting components such as parsers, optimizers, and execution engines through crafted SQL inputs. Depending on the bug types targeted—ranging from crash vulnerabilities~\cite{25sqlsmith,58liu2018treegan,27zhong2020squirrel,28wang2021industry,29fu2022griffin,26jiang2023dynsql,30liang2023sequence,56fu2024sedar} to logic inconsistencies~\cite{08rigger2020testing,12rigger2020findingbugs,11rigger2020detecting,22liang2022detecting,15ba2023testing,18jiang2024detecting,85zhang2025constant} and performance regressions~\cite{43jung2019apollo,44liu2022automatic,42zhang2022learnedsqlgen,45ba2024cert}—existing DBMS fuzzers follow either generation-based or mutation-based strategies.

\begin{sloppypar}
	Generation-based approaches rely on hand-crafted grammars or learned rules to synthesize SQL sequences from scratch. SQLSmith~\cite{25sqlsmith} uses randomized AST construction to generate structurally valid SQL inputs, but often fails on semantic soundness. To enhance expressiveness, DynSQL~\cite{26jiang2023dynsql} dynamically tracks schema evolution to guide generation, while TreeGAN~\cite{58liu2018treegan} adopts grammar-aware GANs to synthesize syntactically valid SQL trees. SQLancer~\cite{08rigger2020testing,12rigger2020findingbugs,11rigger2020detecting,85zhang2025constant} and EET~\cite{18jiang2024detecting} generate ASTs while preserving contextual variable bindings to ensure validity. However, most generators still rely heavily on manual rule engineering or specifications.
	The concurrent work ShQveL~\cite{108zhong2025ShQveL} uses LLM to construct fragments and then uses these fragments to combine SQL statements. This test case generation method still relies on the implementation of general syntax rules.
\end{sloppypar}

\begin{sloppypar}
	Mutation-based fuzzers, in contrast, transform existing SQL statements through changes at the syntax or intermediate representation (IR) level. Some prior work also explores context-dependent mutations arising from the composition of SQL statement sequences.
	Squirrel~\cite{27zhong2020squirrel} represents SQL as intermediate representation (IR) and applies typed, data-aware mutation. RATEL~\cite{28wang2021industry} adds fine-grained coverage tracking and feedback-guided deduplication for robust crash reporting. Griffin~\cite{29fu2022griffin} shuffles SQL statement sequences while tracking metadata dependencies to enable semantically consistent recombination. LEGO~\cite{30liang2023sequence} introduces a type affinity model to mutate SQL type sequences and enrich structural diversity, though semantic restoration remains limited. BUZZBEE~\cite{47yang2024towards} generalizes mutation by abstracting all database interactions into define–use–invalidate triples and uses a lightweight query language (CQL) to ensure semantic compatibility.
\end{sloppypar}

Unlike prior DBMS fuzzers that rely on rigid rule coding or handcrafted dependency models, \toolname\ leverages LLMs to infer semantic dependencies and supports grammar-guided test case synthesis, targeting feature-rich, stateful DBMS.

\subsection{Grammar-Based Fuzzing}
Grammar-based fuzzing generates syntactically valid test inputs by leveraging formal grammars such as context-free grammars (CFGs), addressing the limitations of traditional mutation-based fuzzers that often fail in structured input formats. Early tools like LangFuzz~\cite{77holler2012fuzzing}, CSmith~\cite{67yang2011finding}, NAUTILUS~\cite{70aschermann2019nautilus},and Superion~\cite{78wang2019superion} and FuzzFlow~\cite{107xu2024fuzzing} established grammar-guided mutation or derivation-based input synthesis to maintain structural correctness. Later works incorporated context-sensitivity and probabilistic modeling to capture semantic constraints more effectively, such as Skyfire~\cite{65wang2017skyfire},  SAGE~\cite{66zhou2023towards}, and GrayC~\cite{72even2023grayc}, which learn constraints from corpora or feedback. Other systems like SYNTHFUZZ~\cite{73limpanukorn2025fuzzing}, FreeDom~\cite{74xu2020freedom}, and MLIRSmith~\cite{71wang2023mlirsmith} encode def-use chains, type consistency, or intermediate representations to preserve semantic validity.

Despite these advances, many grammar-based fuzzers still struggle with expressing inter-statement dependencies, environmental assumptions, and stateful execution paths—factors critical in domains like database fuzzing. Some systems mitigate this via probabilistic grammars~\cite{68zhang2024resolverfuzz}, derivation tree splicing~\cite{75alsaeed2023finding}, or grammar automata~\cite{79srivastava2021gramatron}, but semantic gaps remain.
Tribble~\cite{83havrikov2019systematically} systematically explores derivation paths to maximize rule coverage. 
IR-based systems such as POLYGLOT~\cite{69chen2021one}, MLIRod~\cite{76suo2024fuzzing}, and NNSmith~\cite{84liu2023nnsmith} aim to generalize semantics, while Grimoire~\cite{80blazytko2019grimoire}, Evogram~\cite{82gopinath2021input} and Grammar-based Whitebox Fuzzing\cite{81godefroid2008grammar} explore structure without formal grammars. However, integrating semantics into generation pipelines remains complex.

While traditional grammar-based fuzzers emphasize syntactic validity, \toolname\ synthesizes diverse grammar-derived templates and then uses LLMs to force context-aware completion, enabling the generation of corner cases that uncover functionality-specific bugs beyond the expressive reach of fixed grammar rules.

\subsection{LLM-Assisted Fuzzing}
LLMs have recently enabled a new generation of fuzzing tools capable of generating semantically rich and structurally valid inputs across a wide range of domains. For example, CovRL-Fuzz~\cite{88eom2024fuzzing} combines reinforcement learning with LLM-based mutation to improve coverage in JavaScript engines. PromptFuzz~\cite{89lyu2024prompt} mutates prompts instead of programs to synthesize valid C/C++ fuzz drivers guided by runtime feedback. Fuzz4All~\cite{90xia2024fuzz4all} introduces automatic prompting to fuzz heterogeneous systems without manual seed or grammar engineering. In protocol fuzzing, LLMIF~\cite{91wang2024llmif}, CHATAFL~\cite{92meng2024large}, and mGPTFuzz~\cite{102ma2024one} extract protocol structure from natural language documents to guide specification-aware input generation. TitanFuzz~\cite{95deng2023large}, FuzzGPT~\cite{93deng2024large}, KernelGPT~\cite{98yang2025kernelgpt}, and ProphetFuzz~\cite{97wang2024prophetfuzz} target deep learning frameworks and OS kernels by mining vulnerable code patterns or system configurations using LLM reasoning.

Additional efforts explore LLMs as intelligent mutators or input generators. MetaMut~\cite{96ou2024mutators} synthesizes compiler-level mutators from AST APIs and domain prompts. Magneto~\cite{100zhou2024magneto} leverages LLMs to reconstruct multi-stage call chains for exploit generation. ECG~\cite{94zhang2024ecg} uses LLMs to extract syscall specifications and guide semantic mutations in embedded OSs, while CodaMosa~\cite{103lemieux2023codamosa} mixes search-based testing with Codex-based test generation. EAGLEYE~\cite{104liu2025eagleye} targets hidden interfaces in IoT firmware by extracting routing patterns, and InputBlaster~\cite{105liu2024testing} simulates edge-case input behavior in mobile apps via constraint-violating prompt synthesis.

Some studies focus on improving seed quality and diversity~\cite{86oliinyk2024fuzzing}. Sedar~\cite{56fu2024sedar} leverages LLMs for SQL seed adaptation across DBMS dialects but suffers from context hallucination and dependency misalignment.

Notably, \toolname\ does not rely on LLM for mutation, but instead leverages them selectively for semantic instantiation and error-guided repair. 
This integration enables exploration of deeper feature-related paths that remain out of reach for most prompt-only or grammar-guided LLM fuzzers.

\bibliographystyle{ACM-Reference-Format}
%\bibliography{sample-base}
\bibliography{reference}

%%
%% If your work has an appendix, this is the place to put it.
\appendix

\section{Prompts Used In \toolname}

\label{appendix:prompt}

This section documents the exact prompts used in \toolname\ for SQL instantiation and semantic-aware repair.
\begin{promptbox}{SQL Instantiation Prompt}
	Here we initialize a database \texttt{test\_db} by executing the initialization SQL statements:
	\begin{verbatim}
		{init_schema_statements}
	\end{verbatim}
	
	Please instantiate the following SQL statements using the guidance of the given SQL templates:
	\begin{verbatim}
		{sql_templates}
	\end{verbatim}
	
	The generated SQL statements must satisfy the following requirements:
	\begin{itemize}[leftmargin=1.5em, topsep=0pt, itemsep=0pt, parsep=0pt]
		\item They must be syntactically and semantically correct.
		\item They must be executable in \texttt{\{targetDB\}}.
		\item Each subsequent statement must reference only objects created by previous statements (e.g., tables, views, or columns).
		
	\end{itemize}
	
	You are allowed to complete the necessary parts of the templates.
	\textbf{Return only the SQL statements generated from the templates.}
	Do not repeat the initialization SQL statements or include any other content.
	
	Before output, correct any syntax or semantic errors in the generated SQL statements.
	Output the result in JSON format.
	
	\textbf{Example response:}
	\begin{verbatim}
		["SQL1;", "SQL2;", ...]
	\end{verbatim}
\end{promptbox}

\begin{promptbox}{Semantic-Aware Repair Prompt}

	We need you to fix a SQL test case that contains erroneous SQL statements.
	
	The input test case contains multiple SQL statements. SQL statements that require repair are explicitly marked using the following format:
	
	\smallskip
	\begin{verbatim}
		-- [Need to repair<
		<SQL statement>
		-- <error message>
		-- (repair suggestion)
		-- >Need to repair]
	\end{verbatim}
	
	Here is the input test case:
	\begin{verbatim}
		{casecontent}
	\end{verbatim}
	
	You should fix each erroneous SQL statement using:
	\begin{itemize}[leftmargin=1.5em, topsep=0pt, itemsep=0pt, parsep=0pt]
		\item the surrounding SQL context,
		\item the error message, and
		\item the repair suggestion if provided.
	\end{itemize}
	
	\textbf{Target database:} \texttt{\{targetDB\}}.
	
	Do not output anything other than SQL statements.
	Output the fully fixed test case in JSON format.
	
	\textbf{Example response:}
	\begin{verbatim}
		["SQL1;", "SQL2;", ...]
	\end{verbatim}
\end{promptbox}

\section{Vulnerability Status Across DBMSs}
\label{appendix:vuln_status}
This section summarizes the vulnerabilities discovered across multiple DBMSs, and reports their affected components and current handling status (e.g., fixed, verified, or in progress). It provides an at-a-glance view of our disclosure and validation outcomes.

\begin{table}[htbp]
	\scriptsize
	\caption{Vulnerability Discovered Across DBMSs}
	\label{tab:vuln_status}
	\begin{tabularx}{0.75\textwidth}{ c c X X c c }
		
		\toprule
		
		\textbf{DBMS} & \textbf{ID} & \textbf{Reference} & \textbf{Location} & \textbf{Feature-related} & \textbf{Status} \\ \midrule
		\multirow{15}{*}{MySQL} 
		& 1  & CVE-{redacted} & Optimizer & \frNo  & Fixed \\
		& 2  & CVE-{redacted} & Optimizer & \frNo  & Fixed \\
		& 3  & CVE-{redacted} & Optimizer & \frNo  & Fixed \\
		& 4  & CVE-{redacted} & Optimizer & \frNo  & Fixed \\
		& 5  & CVE-{redacted} & Optimizer & \frNo  & Fixed \\
		& 6  & CVE-{redacted} & Optimizer & \frNo  & Fixed \\
		& 7  & CVE-{redacted} & Optimizer & \frNo  & Fixed \\
		& 8  & CVE-{redacted} & Components Services & \frYes & Fixed \\
		& 9  & CVE-{redacted} & Optimizer & \frNo  & Fixed \\
		& 10  & S\{redacted\}     & Item\_ref & \frNo  & Fixed \\ 
		& 11 &  S\{redacted\}       & Tmp\_table & \frNo  & Verified \\ 
		
		& 12 &  S\{redacted\}        & Sql\_executor & \frYes & Fixed \\ 
		& 13 &  S\{redacted\}     & Table\_ref & \frNo  & Fixed \\ 
		& 14 &  S\{redacted\}       & Item\_tree & \frYes & Verified \\ 
		& 15 &  S\{redacted\}        & Mutex & \frYes & Verified \\ 
		& 16 &  S\{redacted\}        & Log\_message & \frYes & Verified \\ 

		& 17 & Debug Build    & Mysql\_execute\_command & \frYes & Verified \\ 
		& 18 & Debug Build    & MDL\_checker & \frYes & Verified \\ 
		
		& 19 & Debug Build    & Explain\_query & \frYes & Verified \\ 
		
		& 20 & Debug Build    & Sp\_head & \frYes & Verified \\ 
		& 21 & Debug Build    & Item\_in\_subselect & \frNo  & Verified \\ 
		& 22 & Debug Build    & User\_table & \frYes & Verified \\ 
		\midrule

		\multirow{12}{*}{MariaDB} 
		&  1 & MDEV-{redacted} & server & \frYes & Verified \\ 
		& 2  & MDEV-{redacted} & Optimizer & \frNo & Fixed \\
		&  3 & MDEV-{redacted} & Optimizer & \frNo & Fixed \\		
		& 4  & MDEV-{redacted} & Temporal Types & \frNo & Verified \\
		&  5 & MDEV-{redacted} & Server & \frYes & Fixed \\ 
		& 6  & MDEV-{redacted} & Server & \frYes & Verified \\ 
		& 7  & MDEV-{redacted} & Server & \frNo & Fixed \\		
		& 8  & MDEV-{redacted} & Optimizer & \frNo & Verified \\
		& 9  & MDEV-{redacted} & Optimizer & \frNo & Fixed \\
		& 10  & MDEV-{redacted} & Optimizer & \frNo & Verified \\
		&  11 & MDEV-{redacted} & Optimizer & \frNo & Verified \\
		&  12 & MDEV-{redacted} & Stored routines & \frYes & Verified \\ 
		& 13 & MDEV-{redacted} & Server & \frYes & Verified \\ 
		&  14 & MDEV-{redacted} & Server & \frYes & Fixed \\ 
		&  15 & MDEV-{redacted} & server & \frYes & Fixed \\ 
		&  16 & MDEV-{redacted} & Optimizer & \frNo & Verified \\
		&  17 & MDEV-{redacted} & Optimizer & \frNo & Verified \\
		&  18 & MDEV-{redacted} & Optimizer & \frNo & Verified \\
		&  19 & MDEV-{redacted} & Server & \frYes & Fixed \\
		&  20 & MDEV-{redacted} & Optimizer & \frNo & Verified \\
		&  21 & MDEV-{redacted} & Server & \frNo & Fixed \\
		&  22 & MDEV-{redacted} & Data types & \frNo & Verified \\		
		&  23 & MDEV-{redacted} & Stored routines & \frYes & Fixed \\		
		&  24 & MDEV-{redacted} & Optimizer & \frNo & Verified \\
		&  25 & MDEV-{redacted} & Add\_key\_field & \frNo & In progerss \\ %bug6
		&  26 & MDEV-{redacted} & Item\_direct\_view\_ref & \frNo & In progerss \\ %bug8
		&  27 & MDEV-{redacted} & Item\_func\_or\_sum & \frNo & In progerss \\ %bug10
		&  28 & MDEV-{redacted} & Setup\_copy\_fields & \frNo & In progerss \\ %bug11
		\midrule
		
		\multirow{2}{*}{SQLite} 
		& 1 & \{redacted\}  & WhereEnd & \frYes & Fixed \\ 
		& 2 & \{redacted\}  & VdbeExec & \frNo & Fixed \\ 
		\midrule

		\multirow{8}{*}{ClickHouse} 
		
		& 1 & \{redacted\}   & FunctionIn & \frNo & Verified \\ 
		& 2 & \{redacted\}   & FunctionBinaryArithmetic & \frNo & Fixed \\ 
		& 3 & \{redacted\}  & InterpreterCreateQuery & \frNo &  Verified \\
		& 4 & \{redacted\}   & Modify  & \frYes & Fixed \\ 
		& 5 & \{redacted\}  & ExecuteQueryImpl & \frYes & Verified \\ 
		& 6 & \{redacted\}   & Planner & \frYes & Fixed \\  
		& 7 & \{redacted\}   & GetColumnName & \frNo & Fixed \\ 
		& 8 & \{redacted\}   & setAlias & \frYes & Verified \\ 
		& 9 & \{redacted\}   & IdentifierResolveScope & \frYes & Fixed \\ 
		& 10 & \{redacted\}   & ExpressionActions & \frNo & Fixed \\ 
		& 11 & \{redacted\}   & Rename & \frYes & Verified \\ 
		& 12 & \{redacted\}   & AssertTypeEquality & \frYes & Fixed \\ 
		
		\bottomrule
		
	\end{tabularx}
\end{table}

\noindent\textbf{Status legend.}
We label an entry as \emph{Fixed} once a patch is available in an upstream release; \emph{Verified} indicates that we reproduced the issue and reported it, but an official fix is still pending; and \emph{In progress} denotes that triage with the maintainers is ongoing. Overall, Table~\ref{tab:vuln_status} summarizes 64 issues across four DBMSs (MySQL: 22,  MariaDB: 28, SQLite: 2, ClickHouse: 12), with identifiers redacted where required by responsible disclosure.

%
%\section{Research Methods}
%
%\subsection{Part One}
%
%Lorem ipsum dolor sit amet, consectetur adipiscing elit. Morbi
%malesuada, quam in pulvinar varius, metus nunc fermentum urna, id
%sollicitudin purus odio sit amet enim. Aliquam ullamcorper eu ipsum
%vel mollis. Curabitur quis dictum nisl. Phasellus vel semper risus, et
%lacinia dolor. Integer ultricies commodo sem nec semper.
%
%\subsection{Part Two}
%
%Etiam commodo feugiat nisl pulvinar pellentesque. Etiam auctor sodales
%ligula, non varius nibh pulvinar semper. Suspendisse nec lectus non
%ipsum convallis congue hendrerit vitae sapien. Donec at laoreet
%eros. Vivamus non purus placerat, scelerisque diam eu, cursus
%ante. Etiam aliquam tortor auctor efficitur mattis.
%
%\section{Online Resources}
%
%Nam id fermentum dui. Suspendisse sagittis tortor a nulla mollis, in
%pulvinar ex pretium. Sed interdum orci quis metus euismod, et sagittis
%enim maximus. Vestibulum gravida massa ut felis suscipit
%congue. Quisque mattis elit a risus ultrices commodo venenatis eget
%dui. Etiam sagittis eleifend elementum.
%
%Nam interdum magna at lectus dignissim, ac dignissim lorem
%rhoncus. Maecenas eu arcu ac neque placerat aliquam. Nunc pulvinar
%massa et mattis lacinia.

\end{document}